\documentclass[useAMS,usenatbib]{mn2e}

\usepackage{graphicx}
\usepackage{url}
\usepackage{float}
\usepackage{color}
\usepackage{soul}

\usepackage{txfonts}
\title[The physical structure of PNe around sdO stars: Abell\,36, DeHt\,2, and RWT\,152]{The physical structure of 
planetary nebulae around sdO stars: Abell\,36, DeHt\,2, and RWT\,152\thanks{Based on 
observations collected at the German-Spanish Astronomical Center, Calar Alto,
jointly operated by the Max-Planck-Institut f\"ur 
Astronomie (Heidelberg) and the Instituto de Astrof\'{\i}sica de
Andaluc\'{\i}a (CSIC); at the 
Observatorio Astron\'omico Nacional in the Sierra San Pedro M\'artir (OAN-SPM), Baja California, Mexico; and
 in The Isaac Newton Telescope, which is operated on the island of
  La Palma by the Isaac Newton Group in the Spanish Observatorio de El Roque de 
Los Muchachos of the Instituto de Astrof\'{\i}sica de Canarias.}}
\author[A. Aller et al.]{A. Aller$^{1,2,3}$, L.~F. Miranda$^{4,3}$,
  L. Olgu\'{\i}n$^5$, R.~V\'azquez$^6$, P.~F. Guill\'en$^6$, R. Oreiro$^4$, 
\newauthor 
A. Ulla$^3$, and E. Solano$^{1,2}$\\
$^{1}$Departamento de Astrof\'{\i}sica, Centro de Astrobiolog\'{\i}a (INTA-CSIC), PO BOX 78, E-28691 Villanueva de la Ca\~nada 
(Madrid), Spain\\
$^{2}$Spanish Virtual Observatory\\
$^{3}$Departamento de F\'isica Aplicada, Universidade de Vigo, Campus Lagoas-Marcosende s/n,  E-36310 Vigo, Spain\\
$^{4}$Instituto de Astrof\'{\i}sica de Andaluc\'{\i}a - CSIC, C/ Glorieta de la Astronom\'{i}a s/n, E-18008 Granada, Spain\\
$^{5}$Departamento de Investigaci\'on en F\'{\i}sica, Universidad de Sonora,
Blvd. Rosales Esq. L.D. Colosio, Edif. 3H, 83190 Hermosillo, Son. Mexico\\
$^{6}$Instituto de Astronom\'{\i}a, Universidad Nacional Aut\'onoma de
M\'exico, Apdo. Postal 877, 22800 Ensenada, B.C., Mexico \\
}

\begin{document}

\date{} 

\pagerange{\pageref{firstpage}--\pageref{lastpage}} \pubyear{}

\maketitle

\label{firstpage}

\begin{abstract}
We present narrow-band H$\alpha$ and [O\,{\sc iii}] images, and
high-resolution, long-slit spectra of the planetary nebulae (PNe) Abell\,36, 
DeHt\,2, and RWT\,152 aimed at studying their morphology and internal
kinematics. These data are complemented with intermediate-resolution,
long-slit spectra to describe the spectral properties of the central stars and
nebulae. The morphokinematical analysis shows that Abell\,36 consists of an 
inner spheroid and two bright point-symmetric arcs; DeHt\,2 is elliptical with protruding polar
regions and a bright non-equatorial ring; and RWT\,152 is bipolar. The
formation of Abell\,36 and DeHt\,2 requires several ejection events including
collimated bipolar outflows that probably are younger than and have disrupted the main shell.
The nebular spectra of the three PNe show a high
 excitation and also suggest a possible deficiency in heavy elements in
 DeHt\,2 and RWT\,152. The spectra of
the central stars strongly suggest an sdO nature and their association with PNe points out
that they have most probably evolved through the asymptotic giant branch. We
analyze general properties of the few known sdOs associated to PNe and find
that most of them are relatively or very evolved PNe, show complex
morphologies, host binary central stars, and are located at
relatively high Galactic latitudes. 
\end{abstract}


\section{Introduction}

Planetary nebulae (PNe) represent the last stage of low- and intermediate-
mass stars (0.8 $\le$ $M$/M$_{\odot }$ $\le$ 8) 
before ending their lives as white dwarfs. It is well known that PNe show
varied morphologies that should be closely related to the mass loss 
history of their progenitor stars. According to M\'endez (1991), most of the
central stars (CSs) can be divided in two different groups: the H-rich and the
H-poor CSs. However, within these two
wide groups, different classes can be found as, e.g., PG\,1159, O, 
Wolf-Rayet-type, and hot subdwarfs CSs.

\begin{table*}
\caption{Common names, PN\,G designations, coordinates, and atmospheric
  parameters of the CSPNe for the objects discussed in this work.}            
\label{table:1}      
\centering
%\begin{center}         
\begin{tabular}{lccccccc}   
\hline\hline           
Object & PN\,G & $\alpha$(2000.0) & $\delta$(2000.0) & {\it $\ell$} & {\it b} & {\it T}$_{\rm eff}  {\rm [K]}$ & log \textit{g}  [cm\,s$^{-2}$]\\   
\hline

Abell\,36  & PN\,G\,318.4+41.4   &   $13^{\rmn{h}} 40^{\rmn{m}} 41\fs3$  & $-19\degr 52\arcmin 55\farcs 3$ & 318$\fdg$4 & 41$\fdg$4  &  93\,000$^{b}$ -- 113\,000$^{c}$ & 5.3$^{b}$ -- 5.6$^{c}$\\

DeHt\,2    & PN\,G\,027.6+16.9  &   $17^{\rmn{h}} 41^{\rmn{m}} 40\fs9$  &  $+03\degr 06\arcmin 57\farcs 3$ & 027$\fdg$6  &16$\fdg$9  &  117\,000$^{d}$ & 5.64$^{d}$ \\

RWT\,152    & PN\,G\,219.2+07.5$^{a}$ & $07^{\rmn{h}} 29^{\rmn{m}} 58\fs5$ &  $-02\degr 06\arcmin 37\farcs 5$ & 219$\fdg$2  & 07$\fdg$5 &    45\,000$^{e}$ & 4.5$^{e}$ \\

\hline

\multicolumn{8}{l}{$^{a}$ Designation proposed in this work following the designation for Galactic PNe by Acker et al. (1992). 
$^{b}$Herrero et al. (1990).} \\
 \multicolumn{8}{l}{ $^{c}$Traulsen et al. (2005). $^{d}$Napiwotzki (1999). $^{e}$Ebbets \& Savage (1982)}\\

\end{tabular}

\end{table*}

Hot subdwarf stars are evolved, low-mass ($M$ $\simeq$ 0.5\,M$_{\odot}$) stars
in the extreme horizontal branch (sdBs) or beyond 
(sdOs). Although it is not well understood how these stars are formed, it is generally
accepted that they are immediate progenitors of white dwarfs (see, e.g., Heber
2009; Geier et al 2011; Geier 2013).  This would imply 
the possibility that many hot subdwarfs (specially the sdOs, hotter than their
cousins sdBs) could be surrounded by PNe. However, there are only a 
few cases of confirmed sdOs as CSs and several cases of PNe whose CSs might
be an sdO, although many of them lack a firm classification yet (see Aller et al. 
2013 and references therein). There is scarce information about PN+sdO systems. Several works
have analyzed the sdOs (or ``possible'' sdOs) themselves 
(see, e.g, Heber et al. 1988; M\'endez 1991; Zijlstra 2007; Montez et
al. 2010; Leone et al. 2011) but there are very few studies about their associated PNe (e.g.,
M\'endez et al. 1988, Goldman et al. 2004). Recently, Aller et al. (2013)
discovered PN\,G\,075.9+11.6, a new PN around the binary sdO 2M1931+4324 (Jacoby et 
al. 2012). The analysis of the object revealed an extremely faint,
double--shell PN, possibly deficient in heavy elements, and hinted to complex ejection 
processes in the formation of the nebula. Another well analysed 
  PN+sdO system is Abell\,41, a bipolar PN with a prominent 
equatorial ring, also around a binary sdO (Bruch et al. 2001, Shimanskii
er al. 2008; Jones et al. 2010). Similar analyses of more PN+sdO
systems may provide important clues about the formation and evolution 
of these objects.

Abell\,36 and RWT\,152 are two faint PNe with sdO CSs and neither their
morphology nor their kinematics have previously been 
analyzed in detail. Abell\,36 was discovered by Abell (1966) using the Palomar
Observatory Sky Survey (POSS) plates and was later imaged by Hua \& Kwok 
(1999). Its bright ($B$ $\simeq$ 11.3\,mag) CS was initially
classified as O(H) by Acker et al. (1992) and as sdO by Kilkenny et
al. (1997). Later, it was incorporated to the Subdwarf 
Database\footnote{\url{http://www.ing.iac.es/ds/sddb/}} by {\O}stensen (2006),
the most complete published compilation of hot subdwarfs to date. 
RWT\,152 is also included in this database as an sdO, although it was
previously classified as an O5 star by Chromey (1980). The PN around RWT\,152 
was discovered by Pritchet (1984) who found a slightly elongated nebulosity
after subtracting the image of a nearby star from the image of RWT\,152 itself.
We have also noticed that the CS of DeHt\,2 has similar spectral features
 to those found in sdOs (see Napiwotzki \& Sch\"onberner
1995, their Fig.\,3) and, therefore, we included it in this investigation. DeHt\,2 was
discovered by Dengel et al. (1980) after inspecting the POSS plates. Its CS
was classified as an O-type star by Acker et al. (1992) and as an hybrid-high
luminosity object by Napiwotzki (1999).

In this work, we present narrow-band, optical images and
high-resolution, long-slit spectroscopy of Abell\,36, DeHt\,2, and RWT\,152,  
which allow us, for the first time, to describe in detail their morphology and
internal kinematics. These data are complemented with 
intermediate-resolution, long-slit spectroscopy of the three objects to
describe the nebular and CS spectra. In the case of DeHt\,2, our data allow us
to classify its CS as a very probable sdO. Table\,1 lists a summary of the three objects: 
the common names and PN\,G designations, the equatorial and Galactic
coordinates, together with the atmospheric parameters (effective temperature and 
surface gravity) of their CSs.

\section[]{Observations}

\subsection{Optical imaging}

Narrow-band H$\alpha$, [O\,{\sc iii}], and [N\,{\sc ii}] images of Abell\,36
were obtained with the Mexman filter-wheel at the 0.84\,m telescope on San Pedro M\'artir Observatory
(OAN-SPM). The [N\,{\sc ii}] image was taken on 2013
February 19 with a seeing of $\simeq$ 2.8 arcsec,  
and the [O\,{\sc iii}] and H$\alpha$ images were obtained on 2013 April 7 with
a seeing of $\simeq$ 2.8 arcsec. A Marconi (e2v) CCD with 2048$\times$4612
pixels each of 15\,$\mu$m in size, was used
as detector in both campaigns. A 2$\times$2 binning was employed 
providing a field of view (fov) of $\simeq$ 8.2$\times$18.4\,arcmin$^{2}$ and a
plate scale of 0.468\,arcsec pixel$^{-1}$. 
Total exposure time was 3600\,s in the
[N\,{\sc ii}] filter ($\lambda_{\rm 0}$ = 6585 \AA, FWHM = 10 \AA), 4800\,s in the [O\,{\sc iii}]
filter ($\lambda_{\rm 0}$ = 5009 \AA, FWHM = 52 \AA), and also 4800\,s in the
H$\alpha$ filter ($\lambda_{\rm 0}$ = 6565 \AA, FWHM = 11 \AA).

Narrow-band H$\alpha$ and [O\,{\sc iii}] images of DeHt\,2 were obtained on 2010 August
23 with the Wide Field Camera (WFC) at the 2.5\,m Isaac Newton Telescope (INT) on El Roque 
de Los Muchachos Observatory (La Palma,
Spain). The detector of the WFC consists of four EEV 2kx4k CCDs 
with a plate scale of 0.33\,arcsec pixel$^{-1}$ and a fov of
34$\times$34 arcmin$^{2}$. Total exposure time was 5400\,s in the 
[O\,{\sc iii}] filter ($\lambda_{\rm 0}$ = 5008 \AA, FWHM = 100 \AA), and
3600\,s in the H$\alpha$ filter ($\lambda_{\rm 0}$ = 6568 \AA, FWHM = 95
\AA). We note that the H$\alpha$ filter includes the [N\,{\sc ii}]$\lambda\lambda$6548,6583 emission lines. However, 
these [N\,{\sc ii}] lines are not detected in the nebular spectra of DeHt\,2 (see Section 3.2.3) and, therefore, 
the H$\alpha$ filter registers only the H$\alpha$ emission line. Seeing was $\simeq$1.5\,arcsec.

In the case of RWT\,152, narrow-band H$\alpha$ and [O\,{\sc iii}] images were
obtained on 2010 December 16 with the Calar Alto Faint 
Object Spectrograph (CAFOS) at the 2.2m telescope on Calar Alto
Observatory (Almer\'{i}a,
Spain). A SITe 2k$\times$2k--CCD was used as detector, with a plate 
scale of 0.53\,arcsec pixel$^{-1}$ and a circular fov of 16 arcmin in
diameter. Total exposure time was 900\,s in [O\,{\sc iii}] filter
($\lambda_{\rm 0}$ = 5007 \AA, FWHM = 87 \AA), and 
1900\,s in the H$\alpha$ filter ($\lambda_{\rm 0}$ = 6563 \AA, FWHM = 15
\AA). Seeing was $\simeq$1.5 arcsec.

The images were reduced following standard procedures 
within the {\sc iraf} and {\sc midas} packages.

\begin{figure*}
  \includegraphics[width=1.0\textwidth]{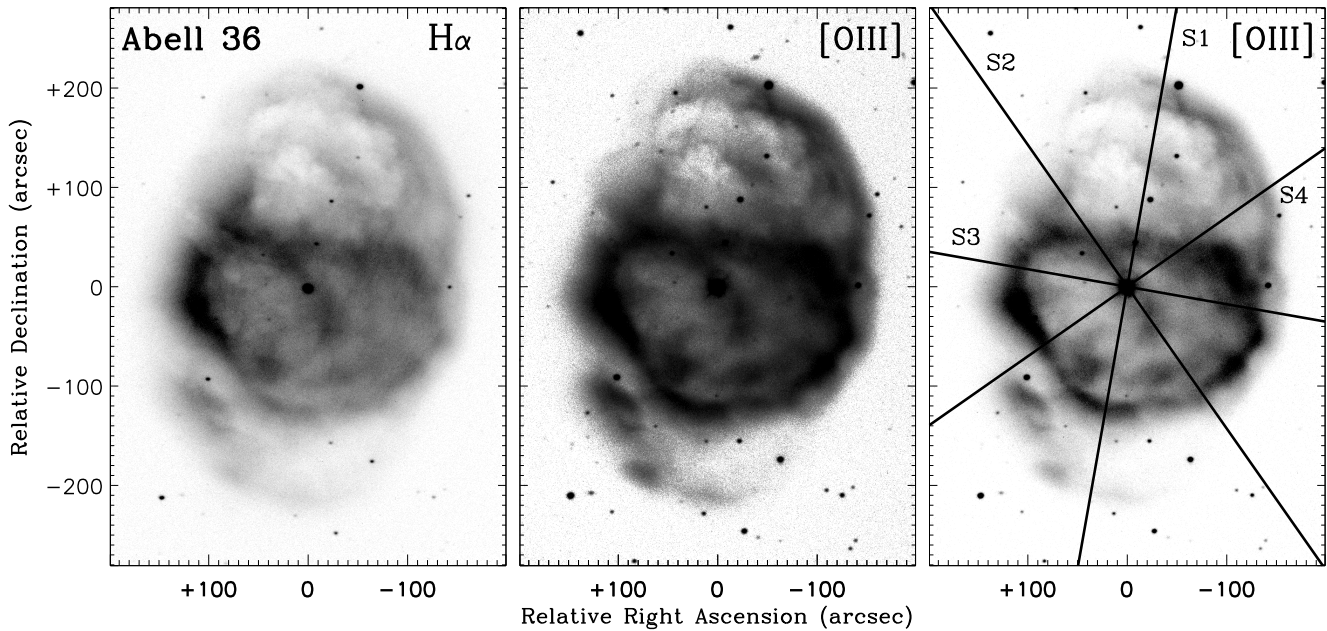}
  \vspace*{1pt}
  \caption{Grey-scale reproductions of the H$\alpha$ (left) and [O\,{\sc iii}]
    (middle and right) images of Abell\,36. Grey levels are linear on 
the left and right panels and logarithmic on the middle one. Slit positions
used for high-resolution, long-slit spectroscopy are drawn in the right panel 
(slit width not to scale).}
\end{figure*}

\subsection{Spectroscopy}

\subsubsection{High-resolution long-slit spectroscopy}

 High-resolution long-slit spectra of Abell\,36, DeHt\,2, and RWT\,152 were
 obtained with the Manchester Echelle Spectrometer 
(MES, Meaburn et al. 2003) at the 2.1\,m telescope on the OAN-SPM during two
different campaigns between 2011 and 2012: Abell\,36 was observed on 2012 May 12-17, DeHt\,2 on 
2012 May 11-13, and RWT\,152 on 2011 February 17. A 2k$\times$2k Marconi CCD
was used as detector in 4$\times$4 
binning (0.702\,arcsec pixel$^{-1}$) in the case of Abell\,36 and DeHt\,2, and
in 2$\times$2 binning (0.338\,arcsec pixel$^{-1}$) 
in the case of RWT\,152. Two filters were used: (1) a $\Delta$$\lambda$ = 60
{\AA } filter to isolate the H$\alpha$ emission line 
(87$^{\rm th}$ order), with a dispersion of 0.11\,$\AA$\,pixel$^{-1}$ (in
4$\times$4 binning) and 0.05\,$\AA$\,pixel$^{-1}$ 
(in 2$\times$2 binning) and (2) a $\Delta$$\lambda$ = 50 {\AA } filter to
isolate the [O\,{\sc iii}] emission line (114$^{\rm th}$
order), with a dispersion 0.08\,$\AA$\,pixel$^{-1}$ (in 4$\times$4 binning) and
0.04\,$\AA$\,pixel$^{-1}$ (in 2$\times$2 binning). Spectra of Abell\,36 and DeHt\,2 were 
obtained with the [O\,{\sc iii}] filter and an exposure time of 1800\,s for each individual spectrum. 
Spectra of RWT\,152 were acquired with the H$\alpha$ and [O\,{\sc iii}] filters and exposures times of 
1200 and 1800\,s, respectively. For all spectra, the slit was centered on the CS of
each PN and oriented at different position angles (PAs) to cover relevant
morphological structures of each object. The observed PAs for each object and
their choice will be described in the corresponding section dedicated 
to each object. The spectra were wavelength calibrated to an accuracy of $\pm$
1 km\,s$^{-1}$ using a Th-Ar lamp. 
The resulting spectral resolution (FWHM) is  12\,km\,s$^{-1}$. Seeing was $\simeq$ 1.5--2\,arcsec 
during the observations.

The spectra were reduced with standard routines for long-slit spectroscopy
within the  {\sc iraf} and {\sc midas} packages. Position-velocity (PV) maps
have been obtained from these high-resolution, long-slit spectra. The origin of 
radial velocities in the PV maps is the systemic velocity obtained for each PN (see
below), and the origin for projected angular distances is the 
position of the CS, as given by the intensity peak of the stellar continuum
that is detected in all long-slit spectra. Internal radial velocities will be quoted
hereafter with respect to the heliocentric systemic velocity of each nebula. The rest 
wavelengths adopted to rescale the radial velocity are 5006.84 $\AA$ for
[O\,{\sc iii}] and 6562.82 $\AA$ for H$\alpha$.

\begin{figure*}
  \includegraphics[width=1.0\textwidth]{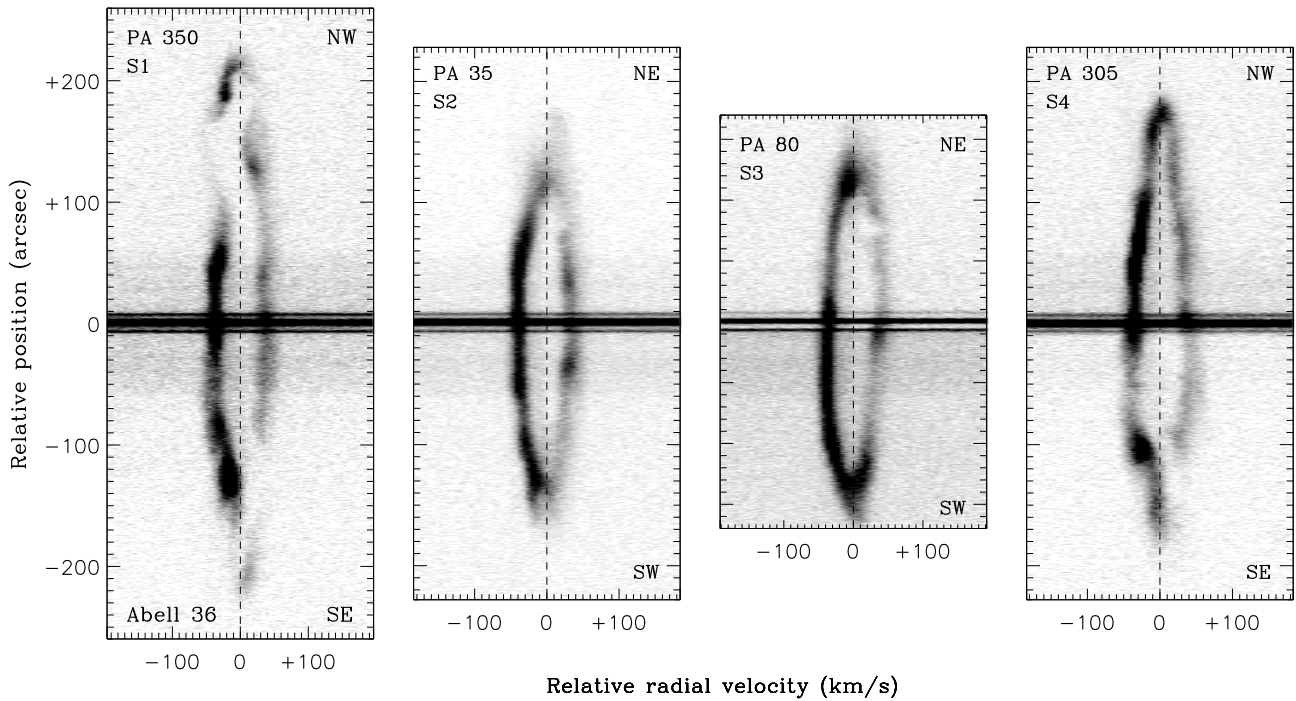}
  \vspace*{10pt}
  \caption{Grey-scale, PV maps derived from the
    high-resolution, long-slit [O\,{\sc iii}] spectra of Abell\,36 
at four different PAs (upper left corner in each panel, see also Fig.\,1). Grey levels are
linear. The origin is at the systemic velocity (see text) 
and position of the CS, as indicated by the stellar continuum. The two horizontal emission features
parallel to the continuum of the CS 
is a well characterized reflection of the instrument.}
\end{figure*}

\subsubsection{Intermediate-resolution, long-slit spectroscopy}

Intermediate-resolution, long-slit spectra of DeHt\,2 and Abell\,36 were obtained with the Boller
\& Chivens spectrograph mounted on the 2.1m 
telescope at the OAN-SPM on 2013 Jun 5 and 6, respectively. The detector was a Marconi CCD with
2k$\times$2k pixels and a plate scale of 1.18\,arcsec\,pixel$^{-1}$. We used a
400 lines\,mm$^{-1}$ dispersion grating, giving a dispersion of
1.7\,$\AA$\,pixel$^{-1}$, and covering the 4100--7600 $\AA$ spectral range. 
In the case of DeHt\,2 a spectrum with 
the slit at PA 55$^{\circ}$ was obtained with the slit centered on the
CS. In the case of Abell\,36 a spectrum with 
the slit at PA 90$^{\circ}$ was obtained covering the CS and the eastern 
part of the nebula. For both objects, the slit width was 2\,arcsec and exposure time was 
1800\,s for each spectrum.  Seeing was $\simeq$ 3\,arcsec.

Intermediate-resolution, long-slit spectra of RWT\,152 were
obtained using CAFOS at the 2.2m telescope on Calar Alto 
Observatory on 2010 December 17. The detector was a SITe 2k$\times$2k--CCD with a plate scale of
0.53\,arcsec\,pixel$^{-1}$. Gratings B-100 and R-100 
were used to cover the 3200--6200 $\AA$ and 5800--9600 $\AA$ spectral ranges,
respectively, both at a dispersion of
$\simeq$ 2\,$\AA$\,pixel$^{-1}$. The spectra were taken with the slit at PA 0$^{\circ}$ and
 exposure times was 900\,s for
each grism. The slit width was 2\,arcsec and it was centered on the
CS. Seeing was $\simeq$ 2\,arcsec. Spectrophotometric standards stars were
observed each night for flux calibration.

The spectra were reduced using 
standard procedures for long-slit spectroscopy within the {\sc iraf} and {\sc
  midas} packages. For each PN, the observed emission line fluxes 
were dereddened using the extinction law of Seaton (1979) and the
corresponding logarithmic extinction coefficient $c$(H$\beta$), as 
obtained from the H$\alpha$/H$\beta$ observed flux ratio, assuming Case B
recombination ($T_{\rm e}$=10$^{4}$\,K, $N_{\rm e}$=10$^{4}$\,cm$^{-3}$) 
and a theoretical H$\alpha$/H$\beta$ ratio of 2.85 (Brocklehurst 1971).

\section{Results}

\subsection{Abell\,36}

\subsubsection{Imaging}

Figure\,1 shows our H$\alpha$ and [O\,{\sc iii}] images of Abell\,36. Our
[N\,{\sc ii}] image does not show nebular emission and is not presented 
here. Abell\,36 presents an elliptical morphology with the major axis oriented
at PA $\simeq$ 350$^{\circ}$, and a size of $\simeq$
7.4$\times$5.3\, arcmin$^2$. Two particularly bright point-symmetric knotty
arcs are observed, giving a spiral appearance to Abell\,36, as already noted by Hua \& Kwok (1999).  
Our images also suggest that a faint elliptical envelope could encircle
the rest of components. The nebular emission is dominated, particularly in [O\,{\sc iii}],  
by a distorted ring-like structure of $\simeq$ 3.3$\times$5.3\,arcmin$^2$ in size, that appears
displaced towards the south with respect to the CS. Several bright knots are
also observed inside this ring. Towards the north, a
bubble-like structure can be recognized inside the elliptical shell, that
apparently emanates from the ring. The bubble extends
up to $\simeq$ 2.6\,arcmin from the CS and is oriented at PA $\simeq$
12$^{\circ}$ that is different from the orientation of the major axis of the
ellipse. 

\subsubsection{High-resolution, long-slit spectroscopy}

Spectra of Abell\,36 were obtained at PAs 35$^{\circ}$, 80$^{\circ}$, 
305$^{\circ}$, and 350$^{\circ}$. The slit positions are 
shown in Fig.\,1 (right panel) overimposed on the [O\,{\sc iii}] image of the nebula, and 
are denoted from S1 to S4 starting at PA 350$^{\circ}$ counterclockwise. The slit PAs were 
chosen to cover the major and minor axis (S1 and S3, respectively) of the ellipse as well as 
two intermediate PAs (S2 and S4). It should be noted that in the cases of PAs
35$^{\circ}$, 305$^{\circ}$, and 350$^{\circ}$, two spectra were secured with the slit on 
the CS but displaced from each other along the corresponding PA
to cover the whole nebula. These two spectra were combined during the
reduction process into a single long-slit spectrum. Figure\,2 shows the PV maps of the 
[O\,{\sc iii}] emission line at the four observed PAs. From the
radial velocity centroid of the [O\,{\sc iii}] emission feature we derive a heliocentric systemic
velocity $V_{\rm HEL}$ = +34.8$\pm$1.4\,km\,s$^{-1}$, in agreement with the value obtained by Bohuski \& 
Smith (1974). 

The PV maps show a velocity ellipse with maximum velocity splitting of $\simeq$ 74\,km\,s$^{-1}$ at 
the stellar position and with no particular tilt with respect to the angular axis. 
The spatio-kinematical properties of the velocity ellipse vary with PA. 
In addition, outer structures are also distinguished. We describe below the PV maps 
in more detail.  

The velocity ellipses at PAs 35$^{\circ}$ and 80$^{\circ}$ (S2 and S3 in Figs.\,1 and 2) present similar 
properties to each other. They extend up to $\simeq$ 130\,arcsec towards the
NE and $\simeq$ 105\,arcsec towards the SW. The size of the velocity ellipses
fits very well the size of the bubble and the southern part of the distorted ring 
(see Fig.\,1), suggesting a spatio-kinematical relationship between both structures.

The PV map at 350$^{\circ}$ (S1 in Fig. 1) reveals a more complex kinematics. The velocity ellipse
extends between $\simeq$ $\pm$ 130\,arcsec from the CS. Two bright knots are observed close to its tips 
with radial velocities of $\simeq$ $\pm$ 20\,km\,s$^{-1}$ (NW knot redshifted). A comparison with the images in 
Fig.\,1 shows that these knots correspond to cuts of slit S1 with the bright edge of the northern bubble 
and with the southern edge of the distorted ring. This result reinforces those obtained at PAs 35$^{\circ}$ and 80$^{\circ}$ 
that the northern bubble and the southern half of the observed ring form an unique spatio-kinematic structure that may be defined 
as a spheroid. It is noteworthy that this spheroid has been identified
  through an analysis of PV maps based on high-resolution spectra and that it can be hardly recognized 
in the direct images. In addition, these spectra also demonstrate that the distorted ring observed in the 
direct images is a projection effect and does not correspond to a real nebular structure. We also note that
the velocity ellipse appears to be open at two point-symmetric
locations on the PV map, with the NW ``hole'' mainly blueshifted and the SE one mainly redshifted. Moreover, 
towards the NW, faint emission, with an arcuate shape in the PV map, and radial velocities up to 
$\simeq$ $-$55\,km\,s$^{-1}$, connects the velocity ellipse with emission from the NW point-symmetric 
arc. The NW arc itself presents two radial velocity components centered at the systemic
velocity, with the brightest component being slightly blueshifted. The emission feature due to the SE arcs is 
similar to that of the NW arcs, but fainter and slightly redshifted. 

At PA 305$^{\circ}$ (S4 in Fig. 1) the velocity ellipse may also be recognized, although it appears open 
at its tips and connected to emission from the point-symmetric arcs. This velocity ellipse is also compatible 
with the spheroidal structures identified at the other PAs. Emission from the NW arc present two velocity 
components, although it is centered close to the systemic velocity. Emission from the SE arc shows a single velocity 
component at the systemic velocity.  

Some of the bright knots observed in the inner nebular regions have been covered by the slits (see Fig.\,1). These knots 
do not appear as separated entities in the PV maps but share the kinematics of the velocity ellipses, suggesting 
that they are a part of the spheroidal structure.

\begin{figure}
  \includegraphics[width=8.8cm]{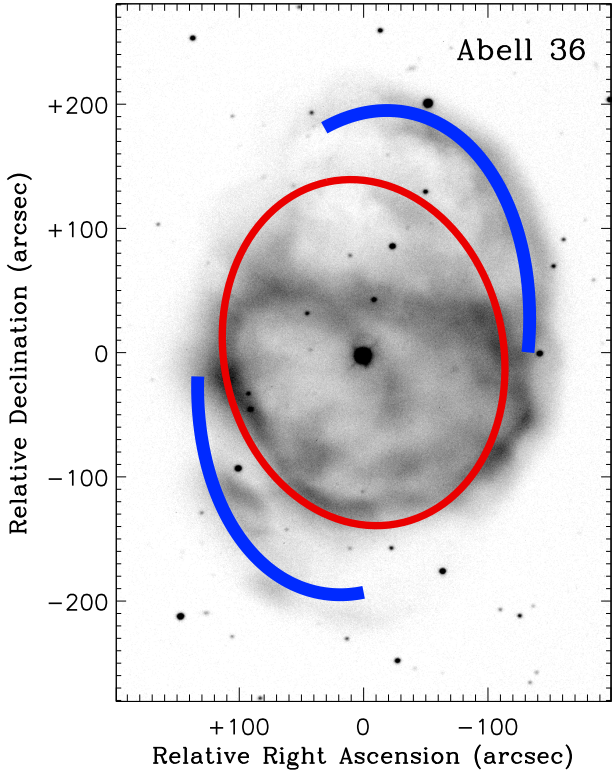}
 \vspace{1pt}
 \caption{Sketch of Abell\,36 as derived from our morphokinematical analysis. The spheroidal shell is displayed 
in red and the bright arcs are represented by thick blue lines 
(see the electronic edition for a color version of this figure).}
\end{figure}

 \begin{table}
\caption{Emission line intensities in Abell\,36.}            
\label{table:1}      
\centering           
\begin{tabular}{lrc}   
\hline\hline           
Line & $f(\lambda)$ & $I(\lambda)$ $I$(H$\beta$)=100) \\   
\hline

H$\gamma$ $\lambda$4340          & 0.129    & 46.5 $\pm$ 0.7 \\

He\,{\sc ii} $\lambda$4686          & 0.042    & 123.0 $\pm$ 1.0 \\

He\,{\sc i}+[Ar\,{\sc iv}] $\lambda$4711          & 0.036    & 16.7 $\pm$ 0.3 \\

[Ar\,{\sc iv}] $\lambda$4740          & 0.029    & 8.4 $\pm$ 0.4 \\

H$\beta$ $\lambda$4861          & 0.000    & 100.0 $\pm$ 1.0 \\

[O\,{\sc iii}] $\lambda$4959    & $-0.023$ &  96.9 $\pm$ 1.0 \\

[O\,{\sc iii}] $\lambda$5007    & $-0.034$ & 278.6 $\pm$ 2.2\\

He\,{\sc ii} $\lambda$5411          & $-0.118$    & 9.5 $\pm$ 0.3 \\

H$\alpha$ $\lambda$6563         & $-0.323$ & 285.0 $\pm$ 1.4 \\

\hline    

$c$(H$\beta$) = 0.17   \\ 

log$F$$_{\rm H\beta}$(erg\,cm$^{-2}$\,s$^{-1}$) = $-$13.34    \\

\hline                                  
\end{tabular}
\end{table}

The PV maps (Fig.\,2) reveal that the [O\,{\sc iii}] emission is noticeable stronger in the 
blueshifted part of the nebula than in the redshifted one, as observed in
other PNe (e.g., IC\,2149, V\'azquez et al 2002). This could be related to
dust absorption of the redshifted emission. Alternatively, interaction of
  the front (blueshifted) half of Abell\,36 with the ISM could be causing
  this effect. To test this possibility, we compare the radial 
  velocity of the local ISM at the position of Abell\,36 with that of the nebula
  itself, assuming a distance of 150--770\,pc (see Abell 1966; Cahn \& Kaler 1971; 
Cahn, Kaler \& Stanghellini 1992; Acker 1998; Phillips 2005) and a standard
rotation curve of the Galaxy. Following the formulation by Nakanishi \& Sofue
(2003), we obtain a heliocentric radial velocity of $\simeq$ +26\,km\,s$^{-1}$
for the ISM around Abell\,36, that is lower than that of the
nebula ($\simeq$ +35\,km\,s$^{-1}$). These values suggest that Abell\,36 is encroaching on the 
ISM, although one would expect that the rear (redshifted) half of Abell\,36 was the brighter 
one, while the opposite is observed. 

The analysis of the PV maps implies a physical structure for Abell\,36 that
is quite different from what could be expected from the images. Figure\,3
shows a sketch of the nebula overimposed on the [O\,{\sc iii}] image. As already mentioned, the velocity ellipse observed at 
all PAs is compatible with a spheroidal structure. Its major axis should be almost 
perpendicular to the line of sight, as indicated by the lack of tilt of the velocity ellipse in the PV maps, and
oriented at PA around 12$^{\circ}$, as suggested by the orientation of the
northern bubble. The arcs resemble the point-symmetric structures observed in other PNe (e.g.,
NGC\,6309, V\'azquez et al. 2008). Hua \& Kwok (1999) compared Abell\,36 with NGC\,6543 and 
our results strength this comparison and extend it to IC\,4364 as well. These
three PNe show an spheroidal/ellipsoidal shell that is accompanied by outer and extended point-symmetric regions 
(components DD' in NGC\,6543 [Miranda \& Solf 1992] and in IC\,4364 [Guerrero et al. 2008] and point-symmetric arcs in Abell\,36), 
which appear twisted with respect to the orientation of the spheroidal
shell. Following these authors, the point-symmetric arcs of Abell\,36 may
be interpreted as due to a collimated bipolar outflow that has been ejected
along a rotating axis. If so, the axis has rotated mainly in a plane (the
plane of the sky) as indicated by the low radial velocity of the arcs, while a
relatively large rotation angle of $\simeq$ 100$^{\circ}$ is inferred from the images.

The velocity ellipses appear
  disrupted at PAs 305$^{\circ}$ and 350$^{\circ}$, where the bright arcs are observed, but not at 
PAs 35$^{\circ}$ and 80$^{\circ}$, where the bright arcs do not extend. This strongly suggests a relationship between 
the bright arcs and the disrupted regions of the spheroid. In
particular, this disruption could be originated by a collimated outflow that
is able to go through the spheroid, perforating parts of it. The
kinematics of the faint emission connecting the velocity ellipse and the
emission features from the arcs observed in the PV map at PA
305$^{\circ}$ strongly suggests an acceleration of material from the 
spheroid followed by a more or less sudden deceleration that could be due to interaction 
with the faint elliptical envelope. It is
worth noting that, if this interpretation is correct, the collimation degree
of the bipolar outflow should have been very high because only ``relatively''
small portions of the spheroid are disrupted at each PA and a velocity
ellipse can still be recognized in the PV maps at PAs 305$^{\circ}$ and
350$^{\circ}$.

The equatorial expansion velocity of the spheroid  
($\sim$ 37\,km\,s$^{-1}$), its equatorial radius ($\sim$ 1.7\,arcmin), 
and the distance (150--770\,pc, see above) yield a kinematical age of 
$\sim$ 2--10$\times$10$^3$\,yr, a broad range of ages given by the
  uncertainty in the distance, that is compatible with a relatively young or very
  evolved PN. Finally, if our interpretation of the bright arcs is correct, the corresponding 
collimated outflows should be younger than the spheroid. However, their
kinematical age is impossible to obtain because their original velocity is
unknown as well as the changes their velocity may have suffered through
collimated outflow--shell interaction.

\subsubsection{Intermediate-resolution, long-slit spectroscopy}

The intermediate-resolution, long-slit nebular spectrum of Abell\,36 is presented in
Figure\,4. A logarithmic extinction coefficient $c$(H$\beta$) of $\simeq$ 0.17 was obtained (see \S\,2.2.2). 
The dereddened line intensities and their Poissonian errors are listed in Table\,2. In addition to 
the hydrogen and [O\,{\sc iii}]$\lambda$$\lambda$4959,5007 emission lines, strong high excitation emission lines 
are observed as He\,{\sc ii}\,$\lambda$4686 and [Ar\,{\sc iv}]$\lambda$$\lambda$4711,4740. We note that 
[O\,{\sc iii}]$\lambda$4363 and [Ar\,{\sc v}]$\lambda$7005 line emissions could also be present but deeper spectra 
are needed to confirm them. The spectrum indicates a high-excitation nebula, which is compatible with the 
non detection of the nebula in the [N\,{\sc ii}] filter.

\begin{figure}
  \includegraphics[width=8.0cm]{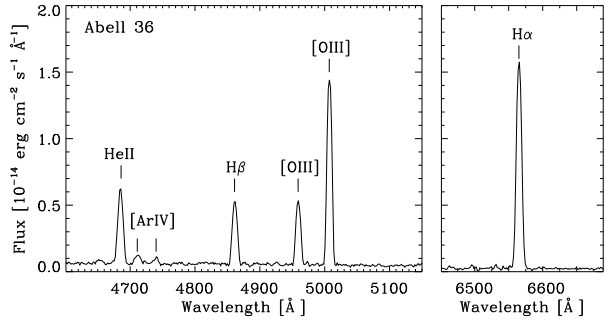}
 \vspace{1pt}
 \caption{Nebular spectrum of Abell\,36 obtained by integrating the emission 
lines between 77 and 124\,arcsec eastern from the CS along the slit position. The 
emission lines are labelled.}
\end{figure}

\begin{figure}
  \includegraphics[width=8.0cm,clip=]{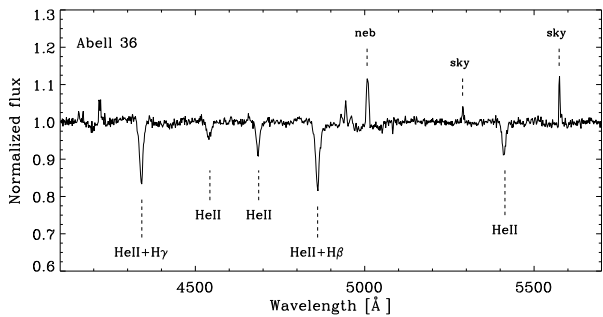}
 \vspace{1pt}
 \caption{Normalized optical spectrum of the CS of Abell\,36. Poorly subtracted sky lines and nebular emission lines are marked as 
well as some absorption lines. }
\end{figure}

The normalized spectrum of the CS of Abell\,36 is shown in Figure\,5. As
already mentioned, this star is included in the Subdwarf
Database by {\O}stensen (2006). The presence of narrow He\,{\sc ii} 
absorption lines (specially He\,{\sc ii}\,$\lambda$4686) as well as the atmospheric
parameters (Table\,1) are indeed compatible with an sdO nature.

\subsection{DeHt\,2}

\subsubsection{Imaging}

Figure\,6 shows our H$\alpha$ and [O\,{\sc iii}] images of DeHt\,2 that reveal
more details than previous ones (Manchado et al. 1996). They show an
elliptical shell with a size of $\simeq$ 1.9$\times$1.5\,arcmin$^2$ and major axis oriented 
at PA $\simeq$ 55$^{\circ}$, although the polar regions seem to protrude and deviate
from a ``pure'' elliptical geometry, in particular at the SW region. The shell
shows a limb-brightening that is more noticeable along the northern edge. This could be a 
result of interaction of the nebula with the interstellar medium (Wareing et al. 2007), an idea that 
is supported by the fact that the limb-brightening is more noticeable in
[O\,{\sc iii}] than in H$\alpha$. Two bright filaments are observed in [O\,{\sc iii}] (much weaker in  H$\alpha$) at the NE
tip of the shell. They are parallel to each other, separated $\simeq$
0.1\,arcmin, and oriented perpendicular to the major nebular
axis. Furthermore, the images reveals the existence of
an (elliptical) ring embedded in the elliptical shell, that is mainly distinguished by its relative brightness. The 
size of the ring is $\simeq$ 1.5$\times$0.7 arcmin$^2$ and its minor axis is oriented E-W
approximately. This ring is drawn in Fig.\,6. The orientations of the ring and the elliptical 
shell are quite different from each other, indicating that the ring does not trace the equatorial 
plane of the elliptical shell.

\begin{figure*}
  \includegraphics[width=1.0\textwidth]{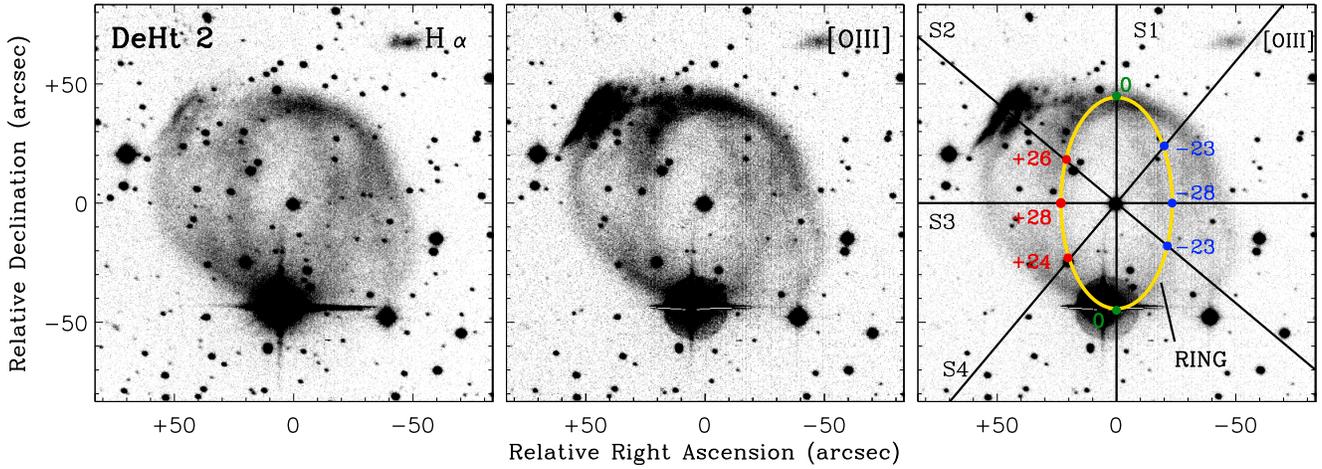}
  \vspace*{1pt}
  \caption{Grey-scale reproductions of the H$\alpha$ (left) and [O\,{\sc
  iii}] (middle and right) images of DeHt\,2. Grey levels are linear. Slit
positions used for the high-resolution, long-slit spectroscopy are drawn on the 
right panel (slit width not to scale). The small nebulosity towards the
northwestern of DeHt\,2 could be a galaxy (see the electronic edition for a color version of this figure).}
\end{figure*}

\begin{figure*}
  \includegraphics[width=1.0\textwidth]{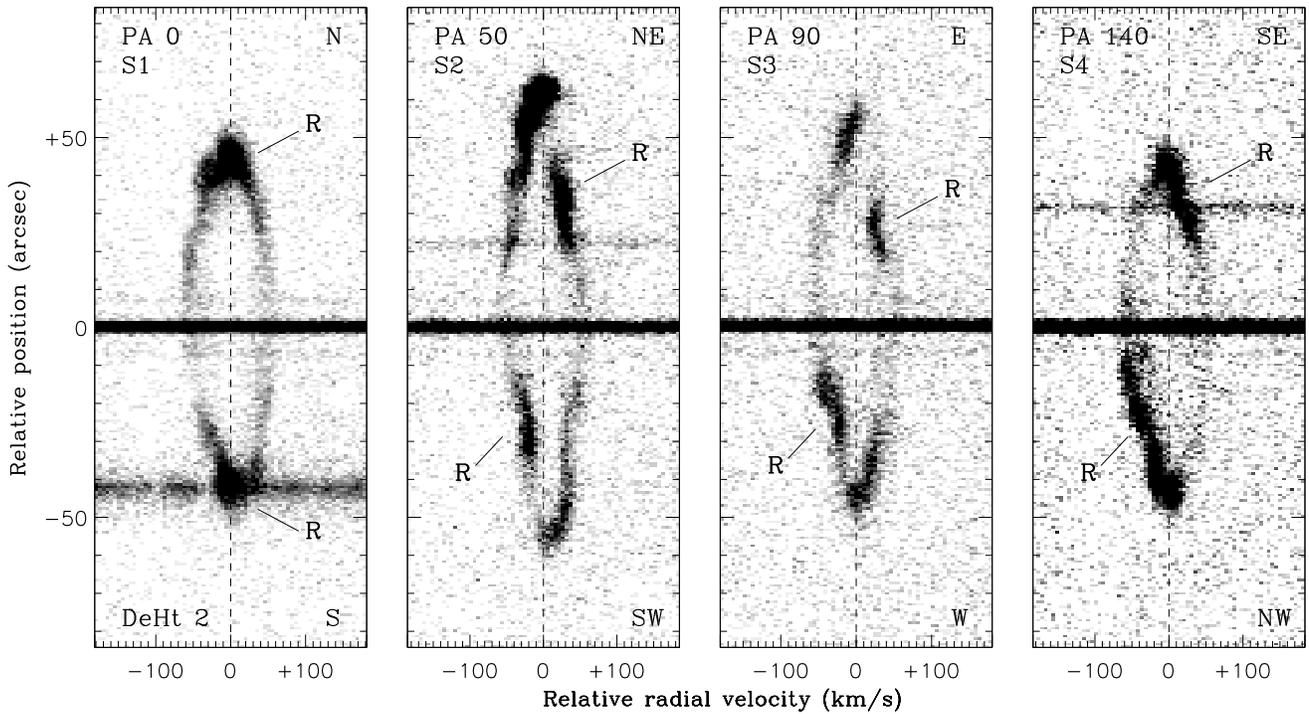}
  \vspace*{10pt}
  \caption{Grey-scale, PV maps derived from the high-resolution, long-slit [O\,{\sc iii}] spectra of DeHt\,2
   at four different PAs (upper left, see also Fig.\,6). The bright features related to the ring-like structure
    are indicated by `R'.}
\end{figure*}

\subsubsection{High-resolution, long-slit spectroscopy}

Spectra of DeHt\,2 were obtained at PAs 0$^{\circ}$,
50$^{\circ}$, 90$^{\circ}$, and 140$^{\circ}$. These slit positions 
(denoted S1 to S4, respectively) are plotted in Fig.\,6 (right panel), on the
[O\,{\sc iii}] image of the nebula. Slits S1 and S3 were chosen to cover the major and 
minor axes of the ring, respectively, while S2 covers the main axis of the elliptical shell and S4 its minor axis. 
Figure\,7 shows the PV maps of the [O\,{\sc iii}] emission line at the four
observed PAs. From the radial velocity centroid of the line emission feature we derive a heliocentric
systemic velocity $V_{\rm HEL}$ = $+$47$\pm$2\,km\,s$^{-1}$. 

The PV maps at PAs 0$^{\circ}$ and 140$^{\circ}$ mainly show a
velocity ellipse. The ellipse does not appear tilted on these PV maps, although
some asymmetries with respect to the velocity axis are observed. At PA
50$^{\circ}$ the emission line feature shows a spindle-like shape slightly tilted in the PV map 
such the NE (SW) regions present an excess of blueshifted (redshifted) radial velocities. 
At PA 90$^{\circ}$ the emission line feature shows a shape halfway between that observed at 
PA 50$^{\circ}$ and PA 140$^{\circ}$. Maximum line splitting of $\simeq$ 100\,km\,s$^{-1}$ is observed at
the stellar position at all PAs. The PV maps also show that the CS is displaced from the nebular centre  
$\simeq$ 3\,arcsec towards the southwest at PA 50$^{\circ}$ and $\simeq$ 5\,arcsec towards the west at PA 90$^{\circ}$, 
which is difficult to recognize in the direct images. By combining these shifts, the CS appears 
displaced $\simeq$ 7.5\,arcsec towards PA $\simeq$ 255$^{\circ}$.

The [O\,{\sc iii}] emission feature is generally weak in the PV maps except at particular
positions that correspond to well identified regions in the
images. The two bright filaments at the NE tip of the elliptical 
shell can be recognized on the PV map at PA 50$^{\circ}$, 
as two knots with radial velocities of $\simeq$ $-$18\,km\,s$^{-1}$ at $\simeq$ 0.9\,arcmin from the 
CS and $-$10\,km\,s$^{-1}$ at $\simeq$ 1\,arcmin. The SW tip of the elliptical shell also appears bright in 
the PV map with a radial velocity of $\simeq$ +16\,km\,s$^{-1}$ at $\simeq$ 0.9\,arcmin. 
The rest of bright regions on PV maps coincide with the ring
identified in the images. Although these features appear elongated in the spatial 
direction (particularly at PA 140$^{\circ}$) and the radial velocity
is difficult to measure, we have considered the position observed in the direct
images to obtain the radial velocity that is indicated in Fig\,6
(right panel). The western half of the ring is blueshifted while the eastern 
half is redshifted. Moreover, the radial velocity presents systematic variations 
in the ring reaching a maximum (in absolute value) of $\simeq$ 28\,km\,s$^{-1}$ at the minor axis, a
minimum value of $\simeq$ 0\,km\,s$^{-1}$ at the major axis, and intermediate
values at PAs 50$^{\circ}$ and 140$^{\circ}$. This kinematics coincides with that expected from a tilted 
circular ring. Under this assumption, we obtain an inclination angle of $\simeq$ 30$^{\circ}$ 
for the plane of the ring with respect to the line of sight, an expansion velocity of $\simeq$ 36\,km\,s$^{-1}$, 
and a PA of $\simeq$ 96$^{\circ}$ for the orientation of the ring axis. It is noteworthy that the
expansion velocity of the ring is lower than the expansion velocity measured 
at the stellar position, $\simeq$ 50\,km\,s$^{-1}$, as indicated by the
maximum radial velocity splitting of the velocity ellipse.

\begin{figure}
  \includegraphics[width=8.8cm]{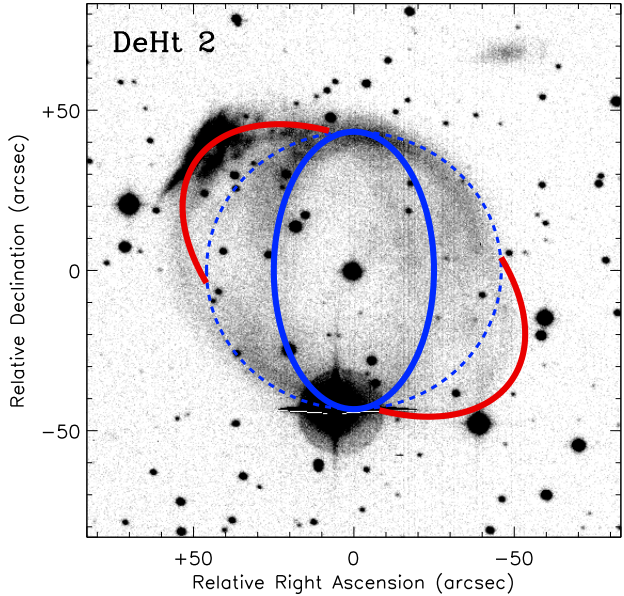}
 \vspace{1pt}
 \caption{Sketch of the  DeHt\,2. as derived from our morphokinematical analysis. The spheroidal shell (dashed blue line), 
 the ring-like structure 
 (solid blue line), and the bipolar outflow (solid red line) are drawn (see the electronic edition for a color version of this figure). }
\end{figure}

The spatio-kinematical properties of DeHt\,2 indicate that its formation has
been complex with at least two different ejection processes being involved. In Figure\,8 we show a schematic representation of 
the structures that compose DeHt\,2, as inferred from the analysis of the PV maps. We suggest that the original 
structure of this PN was a spheroid on which a bright, ring-like region defined its equatorial plane. The fact that
the expansion velocity of the ring (36\,km\,s$^{-1}$) is lower than that measured at the stellar
position (50\,km\,s$^{-1}$) (at some latitude above the equator), strongly suggests that the original 
structure was not spherical but probably an ellipsoid with the major axis oriented approximately
E-W. Taken into account the spatio-kinematical properties of the ring and assuming a distance 
of 1.9--3.2\,kpc (Dengel 1980; Napiwotzki 1999, 2001), its kinematical age results to 
be 1.3--1.9$\times$10$^4$\,yr, compatible with an evolved PN. Probably later,
another bipolar ejection has taken place, that interacted
with and deformed the original spheroid, as suggested by the protruding
regions that are now observed as the polar regions of the apparent elliptical
shell. The second ejection should have been collimated and along a
bipolar axis oriented at PA $\simeq$ 50$^{\circ}$  that is different from the
orientation of the previous structure.

\subsubsection{Intermediate-resolution, long-slit spectroscopy}

Figure\,9 shows the intermediate-resolution, long-slit spectra of the NE filaments of DeHt\,2.
 Only the H$\alpha$, H$\beta$, [OIII]4959,5007, and HeII4686 emission lines are detected. A logarithmic 
 extinction coefficient  $c$(H$\beta$) $\simeq$ 0.33 was obtained (see \S\,2.2.2). The dereddened line
  intensities and their Poissonian errors are listed in Table\,3. 
 The spectrum indicates a very  high excitation although other high-excitation emission lines 
 (as in the case of Abell\,36) are not observed. The same emission lines are detected in other nebular 
 regions (spectra not shown here), suggesting a somewhat lower excitation than in the NE filaments.

 \begin{table}
\caption{Emission line intensities in DeHt\,2.}            
\label{table:1}      
\centering           
\begin{tabular}{lrc}   
\hline\hline           
Line & $f(\lambda)$ & $I(\lambda)$ $I$(H$\beta$)=100) \\   
\hline

He\,{\sc ii} $\lambda$4686          & 0.042    & 103.7 $\pm$ 2.7 \\

H$\beta$ $\lambda$4861          & 0.000    & 100.0 $\pm$ 2.8 \\

[O\,{\sc iii}] $\lambda$4959    & $-0.023$ &  308.5 $\pm$ 3.6 \\

[O\,{\sc iii}] $\lambda$5007    & $-0.034$ & 921.1 $\pm$ 6.5\\

H$\alpha$ $\lambda$6563         & $-0.323$ & 285.0 $\pm$ 4.3 \\

\hline    

$c$(H$\beta$) = 0.33   \\ 

log$F$$_{\rm H\beta}$(erg\,cm$^{-2}$\,s$^{-1}$) = $-$14.63    \\

\hline                                  
\end{tabular}
\end{table}

The stellar spectrum is shown in Figure\,10. It shows strong He\,{\sc ii}
absorptions some of which can be blended with the Balmer absorptions. Although the 
spectrum of DeHt\,2 does not have enough spectral resolution to resolve the
Pickering and Balmer absorption lines, most probably the absorptions present in this
spectrum mainly correspond to the Pickering ones, due to the 
high effective temperature of this CS 
(117000\,K, see Table\,1). These spectral features and the
atmospheric parameters (Table\,1) are compatible with a very hot sdO star. To
provide more support for this classification, we compare in
Figure\,10 the normalized blue spectrum of the CS with that of BD+28$^{\circ}$4211, a well known
sdO with {\it T}$_{\rm eff}$ $\simeq$ 82000\,K and log{\it g} $\simeq$
6.2 cm s$^{-2}$ (Latour et al. 2013). The spectrum of BD+28$^{\circ}$4211 was obtained 
with the CAFOS spectrograph on 2011 July.
 Spectra of both stars are also shown in
Napiwotzki \& Schonberner (1995, their Fig.\,3). Both spectra are remarkably
similar to each other, being the observed differences probably due to the
signal to noise in each spectra and to the different atmospheric parameters of
the stars. In any case, the spectral similarities strongly suggest an sdO nature for the CS
of DeHt\,2.

\begin{figure}
  \includegraphics[width=8.5cm]{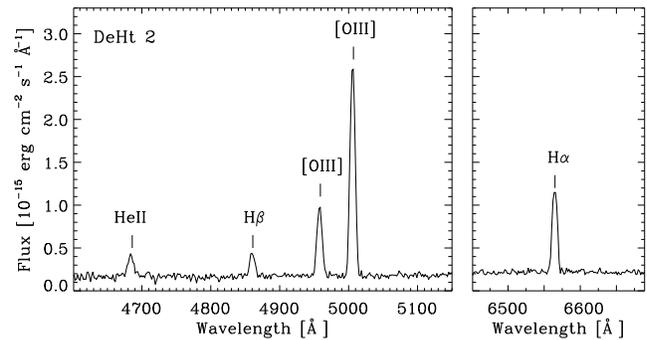}
 \vspace{1pt}
 \caption{Nebular spectrum of DeHt\,2 obtained by integrating the long-slit spectrum between 0.8 and 1.1\,acmin
  from the CS along PA = 55$^{\circ}$. The spectrum corresponds to the 
bright filaments at the northeastern (see Fig.\,6).}
\end{figure}

\begin{figure}
  \includegraphics[width=8.5cm]{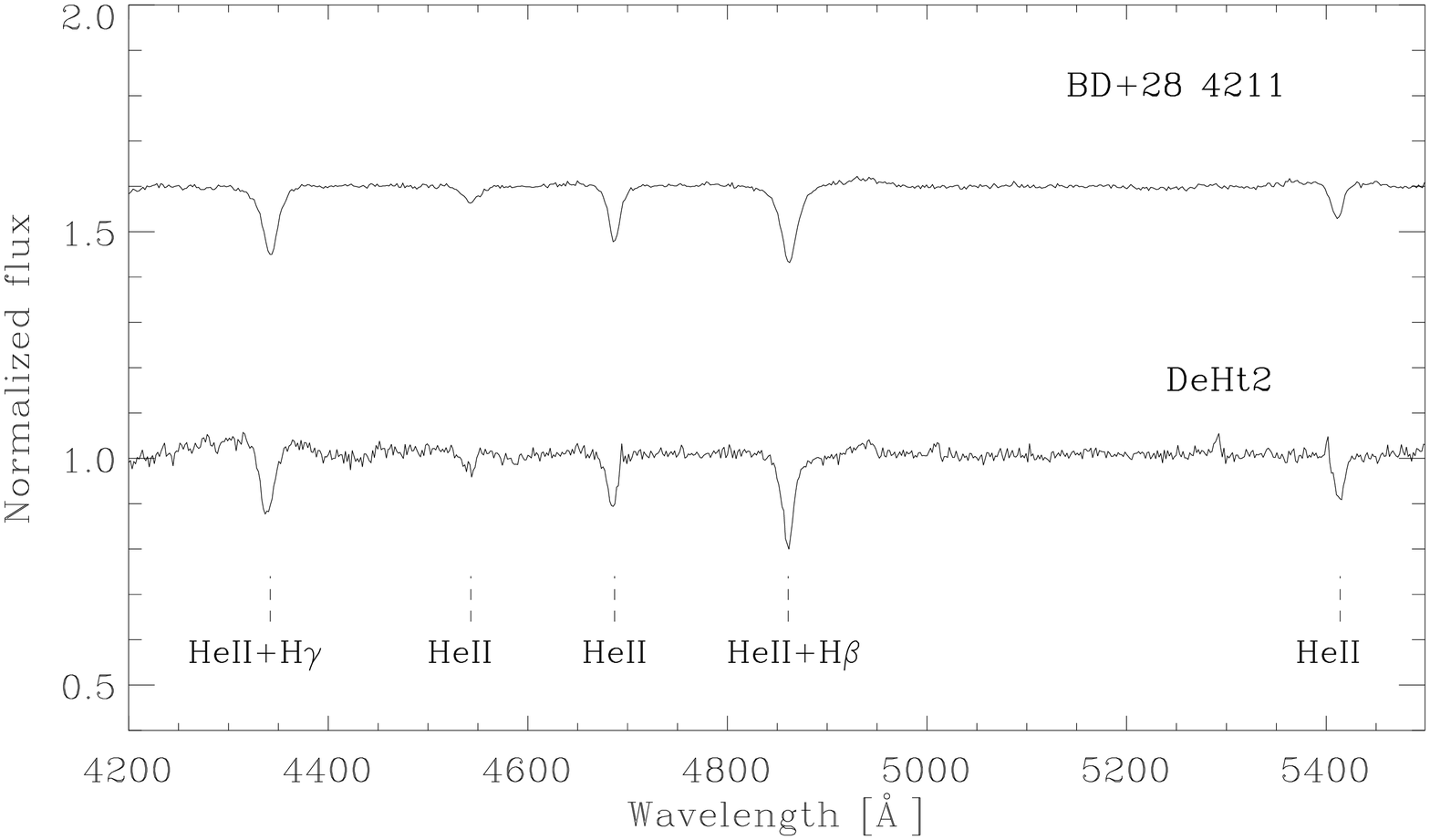}
 \vspace{1pt}
 \caption{Normalized optical spectrum of the CS of DeHt\,2 (bottom) compared to that of 
BD+28$^{\circ}$4211 (top).}
\end{figure}

\subsection{RWT\,152}

\subsubsection{Imaging}

Figure\,11 shows the H$\alpha$ and [O\,{\sc iii}] images of RWT\,152, in which details of 
the nebular morphology can be distinguished for the first time. Both
images reveal a very faint PN. While in the 
H$\alpha$ image the nebula presents a diffuse, although non spherical appearance, a 
more defined nebula can be discerned in the [O\,{\sc iii}] image. At low
intensity levels, the nebula seems to be almost circular 
whereas at higher intensity levels, it appears slightly bipolar
with a size of $\simeq$ 17$\times$21\,arcsec$^2$, major axis oriented 
at PA $\simeq$ 40$^{\circ}$, and a rather uniform intensity distribution. It is worth noting that the 
CS is clearly displaced towards the northwest with respect to the center of the nebula (see also 
below).

\begin{figure*}
  \includegraphics[width=1.0\textwidth]{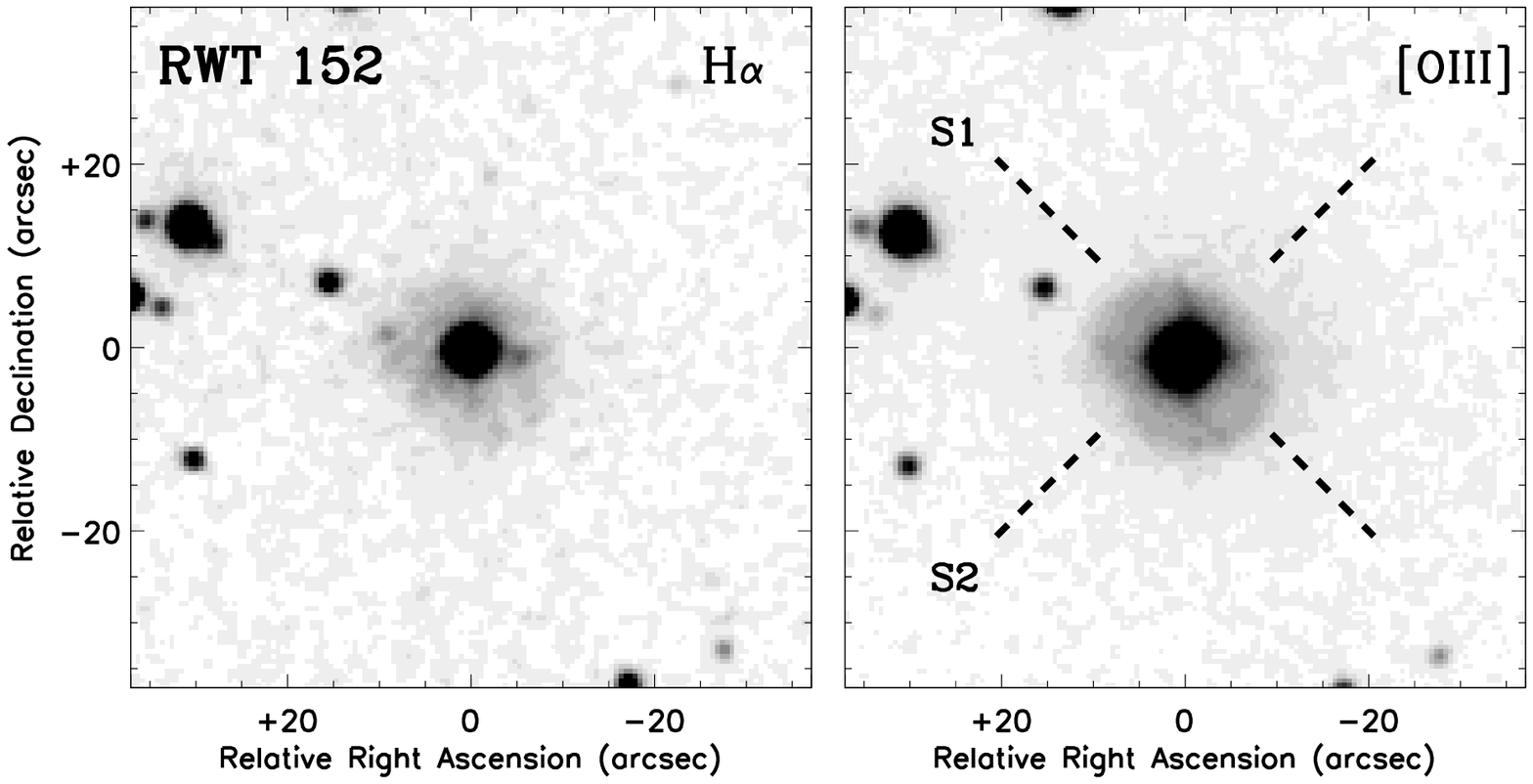}
  \vspace*{1pt}
  \caption{Grey-scale reproductions of the H$\alpha$ and [O\,{\sc iii}] images
    of RWT\,152. Grey-levels are linear. 
A 3$\times$3 box smooth was used for the representation. The slit positions
used for the high-resolution, long-slit spectroscopy 
(S1 and S2) are drawn in the right panel.}
\end{figure*}

\begin{figure*}
  \includegraphics[width=0.65\textwidth]{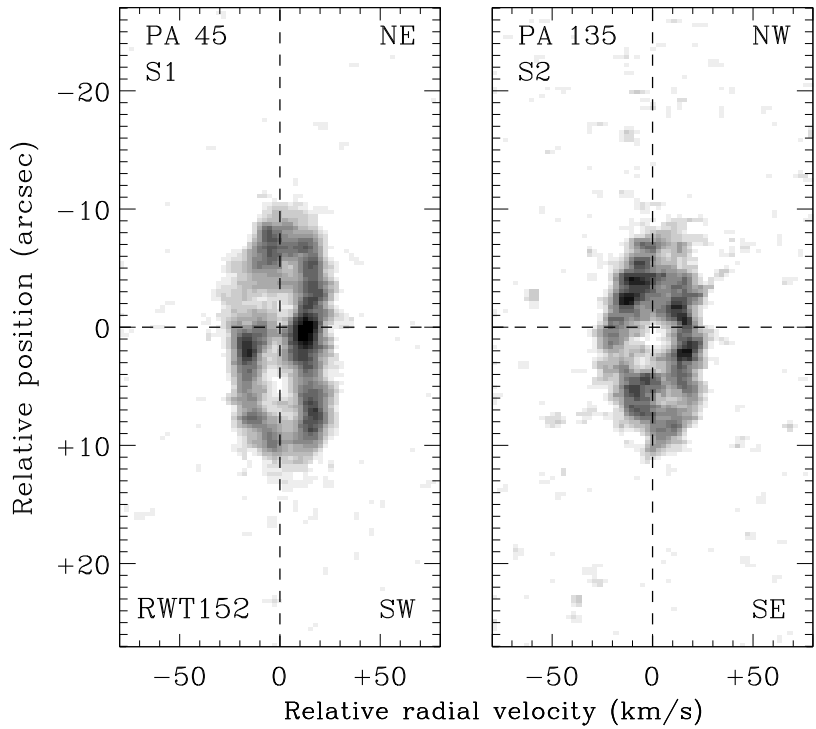}
  \vspace*{10pt}
  \caption{Grey-scale, PV maps derived from the high-resolution, long-slit [O\,{\sc iii}] spectra of RWT\,152. Grey-levels are 
linear. The continuum of the central star has been removed (using the \textit{background} 
{\sc iraf} task) and its position is marked with a dashed horizontal line. A 3$\times$3 box smooth was used for the representation. }
\end{figure*}

\subsubsection{High-resolution, long-slit spectroscopy}

Spectra of RWT\,152 were obtained at PA 45$^{\circ}$ (in
H$\alpha$ and [O\,{\sc iii}]) and PA 135$^{\circ}$ 
(in [O\,{\sc iii}])  to cover the major and minor axis (S1 and S2, respectively, in
Fig.\,11) of the bipolar shell. Figure\,12 shows the two PV maps in the [O\,{\sc iii}] 
emission line. We note that the [O\,{\sc iii}] emission line feature presents a much more knotty
appearance in the PV maps than in the image. The PV map of H$\alpha$ emission line at PA 45$^{\circ}$ (not shown here) is
very similar to that of the [O\,{\sc iii}] emission line at the same
PA. but the large thermal width in the H$\alpha$ line does not allow us a
detailed analysis of the kinematics. From the velocity
centroid of the line emission features we derive a heliocentric systemic velocity 
of $V_{\rm HEL}$ = +134.5$\pm$1.8\,km\,s$^{-1}$. 

In the PV map at PA 45$^{\circ}$, the [O\,{\sc iii}] emission
feature presents an hour-glass like shape with a size of $\simeq$ 22\,arcsec, although deviations 
from a pure hour-glass shape are noticed, particularly in the NE lobe. Radial velocity at the tips of the emission feature is 
$\simeq$ $\pm$ 7\,km\,s$^{-1}$ . Two bright knots can be distinguished in the central region, 
that are symmetric in radial velocity but not centered on the CS: the redshifted knot 
presents a radial velocity of $\simeq$ +14\,km\,s$^{-1}$ and is located $\simeq$ 0.12\,arcsec NE from the
CS; the blueshifted one has a radial velocity of $\simeq$ $-$14\,km\,s$^{-1}$
and is located $\simeq$ 1.8\,arcsec SW from the CS. The PV map at PA 135$^{\circ}$ presents a
velocity ellipse with a maximum line splitting of $\simeq$ 34\,km\,s$^{-1}$ at
the stellar position and a size of $\simeq$ 18\,arcsec as measured between the intensity
peaks at the systemic velocity. The centre of the velocity ellipse is displaced $\simeq$ 1.2\,arcsec 
towards the SE with respect to the CS. 

The displacements of nebula's centre with respect to the
CS as measured in the PV maps are consistent with the off-centre
position of the CS observed in the direct images. Taken into account the two observed
PAs, a shift of $\simeq$ 1.4\,arcsec towards PA $\simeq$ 348$^{\circ}$ is obtained.

Both images and PV maps are compatible with a bipolar PN. The two
bright knots observed in the central regions in the PV map at PA 45$^{\circ}$
suggest the existence of an equatorial enhancement. If we
assume circular cross section for the equator, the equatorial plane of the nebula is tilted 
by $\simeq$6$^{\circ}$ with respect to the line of sight. Assuming homologous expansion, a polar velocity of 
19\,km\,s$^{-1}$ is obtained. There is no reliable determination for the distance of RWT\,152 and estimates 
are 1.4 and 6.5\,kpc (Ebbets \& Savage 1982; Pritchet 1984). In consequence, only a lower limit of 
$\simeq$ 4$\times$10$^3$\,yr can be obtained for its kinematical age, which
suggests (at least) a relatively evolved PN.

\subsubsection{Intermediate-resolution, long-slit spectroscopy}

The intermediate-resolution nebular spectrum of RWT\,152 is presented in
Figure\,13. Only the H$\alpha$, H$\beta$ and [O\,{\sc
iii}]$\lambda$$\lambda$4959,5007 emission lines are identified. A logarithmic extinction coefficient
$c$(H$\beta$) of $\simeq$ 0.46 was derived (see \S\,2.2.2). Table\,4 lists the
dereddened line intensities and their Poissonian
errors. The [O\,{\sc iii}]/H$\beta$ line intensity ratio is $\simeq$ 8 (Table\,4), suggesting a 
relatively high excitation.

The normalized spectrum of the CS is shown in Figure\,14. In contrast to the
CS spectrum of DeHt\,2, the CS spectrum of RWT\,152 is dominated by hydrogen
Balmer lines. The narrowness of the absorption lines, and the presence of 
He\,{\sc i} (e.g. He\,{\sc i} $\lambda$$\lambda$4386,4471) and He\,{\sc ii}
absorption lines (specially He\,{\sc ii} $\lambda$ 4686) confirm the sdO 
nature of the CS. The CS was analyzed by Ebbets \& Savage (1982) who
determined a relatively low (for an sdO) \textit{T}$_{\rm eff}$ of $\simeq$
45000\,K (see Table\,1) that is compatible with the
presence of He\,{\sc i}$\lambda$4471.

\section{Discussion}

The data presented and analyzed in the previous sections have allowed us to deduce the basic physical structure 
and emission properties of Abell\,36, DeHt\,2, and RWT\,152 and their CSs. Moreover, the spatio-kinematical analysis has been able 
to recover relevant information about the processes involved in the formation
of the three objects. In addition, the spectra of the three CSs show characteristics that
allow us to classify them as sdOs. In particular, the narrowness of the 
absorption lines and the presence of prominent He\,{\sc ii} absorption are typical of sdOs. This classification is corroborated 
by the atmospheric parameters of the CSs (Table\,1), that are within the range of the sdOs atmospheric parameters
 (see Heber 2009).

RWT\,152 seems to be a result of a typical bipolar ejection as observed in
many PNe. The formation of Abell\,36 and DeHt\,2 appears more complex and requires multiple
ejection events, changes in the orientation of main ejection axis
between events, and a different collimation degree of the ejections. In Abell\,36 
the bright arcs indicate a very large and ``continuous'' change in the collimated ejection 
axis, whereas in DeHt\,2 the bipolar outflows seem to have acted along a constant direction 
that is different from the main axis of the previous shell. Interestingly,
evidence is found in both PNe that the collimated outflows might have 
been ejected after the main nebular shell was formed. Moreover, in both cases,
the collimated outflows seem to have disrupted or deformed the previous
shell. This situation is similar to that found in other PNe (e.g., Guerrero \&
Miranda 2012; Ramos-Larios et al. 2012; Guill\'en et al. 2013) in which young
collimated outflows seem to have disrupted a previous nebular structures. The
origin of collimated outflows in PNe after the formation of the 
main nebular shell is difficult to explain within current scenarios for PN formation 
and is still matter of debate (see Tocknell, De Marco \& Wardle 2014).

 \begin{table}
\caption{Emission line intensities in RWT\,152.}            
\label{table:1}      
\centering           
\begin{tabular}{lrc}   
\hline\hline           
Line & $f(\lambda)$ & $I(\lambda)$ $I$(H$\beta$)=100) \\   
\hline

H$\beta$ $\lambda$4861          & 0.000    & 100 $\pm$ 3 \\

[O\,{\sc iii}] $\lambda$4959    & $-0.023$ &  209 $\pm$ 3 \\

[O\,{\sc iii}] $\lambda$5007    & $-0.034$ & 591 $\pm$ 3\\

H$\alpha$ $\lambda$6563         & $-0.323$ & 285 $\pm$ 4 \\

\hline    

$c$(H$\beta$) = 0.46   \\ 

log$F$$_{\rm H\beta}$(erg\,cm$^{-2}$\,s$^{-1}$) = $-$14.97    \\

\hline                                  
\end{tabular}
\end{table}

Multiple ejection events, as those identified in Abell\,36 and DeHt\,2, are observed
in many PNe. The idea that complex PNe are related to the evolution of binary
CSs has been present during many years and it has received strong support with
recent detections of new binary CSs in PNe with multiple structures and jets
(see Miszalski et al. 2009 and references therein). Within this context, it could be suggested 
that the CSs of Abell\,36 and DeHt\,2 are also binaries, although, to the best of our knowledge, no direct
evidence exists for such binaries. It is interesting to note that both
Abell\,36 and DeHt\,2 contain off-center CSs, which could be considered as an indirect 
evidence for a binary CS (e.g., Soker et al. 1998). Although it is true that some binary CSs appear 
off-center, inferring a binary CS from its off-center position only should be seen
with caution. Given that DeHt\,2 and, perhaps, Abell\,36 are evolved PNe, the
off-center CSs could be caused by deformation in the shell due to, for
instance, interaction with the interstellar medium (Jones et. al 2010; Frew et al. 2014), 
and/or amplification through evolution of (originally small) asymmetries in the
ejection process. In the case of Abell\,36, interaction shell-collimated
outflows could also contribute to create asymmetries in the shell. The case of
the off-center CS of RWT\,152 looks different
because of the more symmetric shell. However, RWT\,152 may be
a very distant PN and a higher spatial resolution is necessary to investigate
possible asymmetries in the shell, which are already suggested by the
distortions in the kinematics. In any case, these three PNe are good candidates to
host binary CSs (see also below), and dedicated observations of their CSs should be obtained to
search for possible companions. 

The nebular spectra of Abell\,36 indicates high excitation, as shown by the presence of 
[Ar\,{\sc iv}] and strong He\,{\sc ii}\,$\lambda$4686 emission lines. The
nebular spectra of DeHt\,2 and RWT\,152 also indicate high excitation but 
  no emission lines from heavy elements (except  [O\,{\sc iii}]) are
  detected. If other emission lines exist in these two PNe, they should be
very faint. The CSs of Abell\,36 and DeHt\,2 present very similar
atmospheric parameters (Table\,1). Therefore, similar emission lines could be
expected, unless the physical conditions and/or chemical abundances are very
different in both PNe. The CS of RWT\,152 has a relatively low $T$$_{\rm efff}$ and, in
principle, low-excitation emission lines should be present in the
nebula. The nebular spectra of DeHt\,2 and RWT\,152 are very similar 
to that of PN\,G\,075.9+11.6 (Aller et al. 2013), in which only [O\,{\sc iii}] and
Balmer emission lines have been detected. 
Following these authors, a probable explanation 
for the peculiar nebular spectra of DeHt\,2 and RWT\,152 is a deficiency in
heavy elements in the nebula. Such a deficiency may be expected in PNe that originate 
from low-mass progenitors (see, e.g., IC\,2149, V\'azquez et al. 2002) and it would be 
consistent with the idea that sdOs evolve from low-mass progenitors (see, e.g., Heber 2009). Deep spectroscopy of
these PNe is crucial to detect faint emission lines and to obtain their chemical abundances.

 \begin{figure}
  \includegraphics[width=8.5cm]{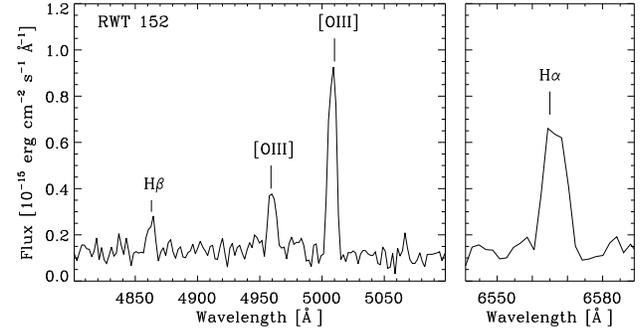}
 \vspace{1pt}
 \caption{Nebular spectrum of RWT\,152 obtained 
by integrating the detected emission lines 4.3 and 8.6 arcsec northern of RWT\,2 along PA = 0$^{\circ}$.}
\end{figure}

\begin{figure}
  \includegraphics[width=8.0cm]{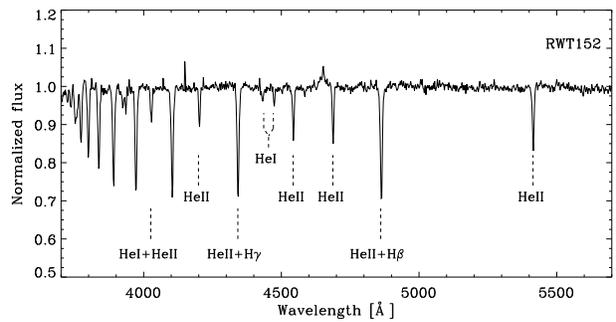}
 \vspace{1pt}
 \caption{Normalized optical spectrum of the central star of RWT\,152. Some of the absorption lines (specially He\,{\sc i} 
and He\,{\sc ii}) are marked.}
\end{figure}

 \begin{table*}
\caption{Properties PN+sdO systems}            
\label{table:1}      
\centering           
\begin{tabular}{lllll}   
\hline

PN\,G & Name & Morphology / Comments & Binary CS & References \\   

\hline

009.6+10.5 & Abell\,41  & Bipolar   & Y & (1), (2), (3)  \\

009.8--07.5 & GJJC\,1   & Irregular / Cometary-like    & ? &  (4),(5) \\

027.6+16.9 & DeHt\,2    & Elliptical / Spheroidal shell and bipolar outflows
at different orientations & ? & (6) \\

053.8--03.0 & Abell\,63 & Bipolar / Jets   & Y & (7),(8),(9), (10)\\

055.4+16.0  & Abell\,46 & Elliptical-Bipolar  & Y & (9),(11),(12),(13) \\

065.0--27.3 & K\,648    & Elliptical / Two elliptical shells and halo & ? &
(14)   \\

075.9+11.6 & 2M1931+4324 & Multishell / Bipolar and elliptical shell at different orientations & Y &  (15),(16) \\

136.3+05.5 & HFG\,1     &  Irregular   & Y &  (17),(18) \\

215.6+03.6  & NGC\,2346  & Bipolar   & Y &  (9),(12),(19) \\

219.2+07.5 & RWT\,152   & Bipolar    & ? & (6),(20)  \\

272.1+12.3 & NGC\,3132  & Elliptical  & Y & (21),(22) \\

273.6+06.1 & LSS\,1362  & Irregular-Elliptical  & N & (23),(24) \\

279.6--03.1 & He\,2-36  & Elliptical / Point-symmetry & ? & (25),(26)  \\

283.9+09.7 & LSS\,2018 (DS\,1)  &  Bipolar-Irregular / Low-ionization structures & Y & (9),(27)\\

318.4+41.4 & Abell\,36 & Elliptical / Spheroidal
shell and point-symmetric arcs  & ? & (6),(20) \\

331.3--12.1 & Hen\,3--1357 & Multishell / Bipolar and elliptical shells and
jets & ? & (28),(29)  \\

335.5+12.4 & LSE\,125 (DS\,2) & Round   & N & (27),(30) \\

339.9+88.4 & LoTr\,5    & Bipolar   & Y &  (31),(32) \\

\hline    
\end{tabular}

(1) Bruch et al. (2001); (2) Shimanskii et al. (2008); (3) Jones et al. (2010)
(4) Borkowsky et al. (1993); (5) Rauch et al. (1998); (6) This work; (7) Pollacco \&
Bell (1997); (8) Tsessevich (1977); (9)
Miszalski et al. (2009); (10) Mitchell et al. (2007); (11) Stanghellini et
al. (2002); (12) Bond \& Livio (1990); (13) Ritter \& Kolb (2003); (14) Alves et al. (2000);
(15) Aller et al. (2013); (16) Jacoby et
al. (2012); (17) Heckathorn et al. (1982); (18) Grauer et al. (1987); 
(19) Kohoutek \& Senkbeil (1973); (20) {\O}stensen (2006); (21) Ciardullo et
al. (1999); (22) Monteiro et al. (2000); (23) Chu
et al. (2009); (24) Heber et al. (1988); (25) Corradi \& Schwarz (1993); (26) M\'endez
(1978); (27) Drilling (1983); (28) Bobrowsky
et al. (1998); (29) Reindl et al. (2014);
(30) Hua et al. (1998); (31) van Winckel et al. (2014); (32) Graham et
al. (2004)

\end{table*}

It is interesting to compare the properties of PN+sdO systems. A search in the
literature suggests 33 PNe with sdO CS, although in several cases this
classification is doubtful or not confirmed (see Weidman \& Gamen
2011). For instance, the CS of NGC\,2371 is classified as sdO in SIMBAD but as
WR-PG\,1159 by Herald \& Bianchi (2004); the CS of NGC\,6026 is classified as
WD/sdO by de Marco (2009), as a pre-WD/WD by Hillwig et al. (2010) but as OB
by Weiddman \& Gamen (2011); in the case of NGC\,1514, the hot star in its 
binary CS has been classified as sdO (Kohoutek 1967) but a recent spectral analysis (Aller
et al. 2014, in preparation) does not allow us to establish a firm classification. If we
restrict to those objects with a more confident classification, the number of
PN+sdO systems is 18. Table\,5 lists these PNe (columns 1
and 2), their morphology and some comments about nebular structures present in
them (column 3), the binary nature of the CSs (column 4), and the corresponding references
(column 5). We emphasize that the discussion below does not critically depends on
whether some of the possible PN+sdO systems are added to the list and/or whether
some of the objects in Table\,5 are removed. We note that our
Table\,5 shares several objects with Table\,1 by Miszalski et
al. (2009). These authors analyze the morphology of PNe with close binary CSs
while we focus in the morphology and presence of binaries in sdO CSs. 

An inspection of the properties of PN+sdO systems reveals that most of these PNe are
very faint, suggesting that they are in a moderately or very evolved stage. A
noticeable exception is the Stingray nebula (Hen\,3$-$1357), a very
young PN (Parthasarathy et al. 1995) whose CS (SAO244567) has recently been
identified as an sdO (Reindl et al. 2014). If we attend to the morphology, the
sample is dominated by elliptical and bipolar shapes with only an object
(DS\,2) presenting a round morphology. Moreover, many of these PNe show
multiple structures, jet-like outflows or point-symmetric structures that
could be related to the action of bipolar collimated outflows. It is also
remarkable that many PN+sdO systems host binary CSs, suggesting that binary
CSs may play an important role in the formation of PN+sdO systems. Finally, a
large fraction of these systems are observed at a relatively high Galactic latitude. In particular, 11
PNe in Table\,1 have $|b|$ $>$ 10$^{\circ}$ and 15 have $|b|$ $>$
7$^{\circ}$. Although the number of objects is small to drawn firm
conclusions, the relatively high Galactic latitudes are more typical of round PNe
evolving from low-mass progenitors (Corradi \& Schwarz 1995;
Stanghellini et al. 2002). Remarkably, low-mass progenitors are generally expected for
sdOs (see above) and, in addition, sdOs are normally located at high Galactic latitudes. However,  
one would not expect a large fraction of complex PNe resulting from low-mass
progenitors. These results and the large fraction of binary CSs in PN+sdO
systems point out that the key parameter to form a
complex PN is a binary CS rather than the mass of the progenitor. This
conclusion is reinforced by recent results by, e.g., Miszalski et al. (2009), Boffin et al. 
(2012), Corradi et al. (2014), and Jones et al. (2014) who found that most PNe with
close binary CS present complex morphologies.

sdOs associated to PNe represent a very small fraction of the
$\geq$ 800 known sdOs ({\O}stensen 2006). This number could increase as more
CSs may be classified as sdO (e.g., Reindl et al. 2014). In this respect, we note the 
lack of firm classifications for many CS of PNe, which are crucial to identify 
new sdO among CSs. On the other hand, a recent image survey of $\simeq$ 80 ``classical'' sdOs 
(Aller et al., in preparation) to search for associated PNe has identified only a new case 
(Aller et al. 2013). Nevertheless, given the intrinsic faintness of these PNe, much 
deeper surveys should be carried out to identify more cases. Several evolutionary paths
are considered for the formation of sdOs (see Heber 2009) and, in the case of
PN+sdO systems, evolution through asymptotic giant branch (AGB) and post-AGB
phases appears to be the most suitable one. For the rest of sdOs (those
without a PN), other evolutionary paths (post-red giant branch, post-extended horizontal branch
evolution, or star mergers) should be considered (Alves et al. 2000; Napiwotzki 2008; Heber 2009). 
Nevertheless, if sdO evolution proceeds very slowly, the AGB/post-AGB
evolution of a fraction of sdOs may go unnoticed if the nebula has dispersed
before being photoionized. In any case, identification of new PN+sdO systems,
and detailed studies of the known ones are important because these systems may
provide clues about the formation of complex PNe and sdO evolution.

\section{Conclusions}

We have presented and analyzed narrow-band direct images, and high- and
low-resolution, long-slit spectra of Abell\,36, DeHt\,2, and RWT\,152, 
three PNe for which detailed spatio-kinematical analysis had not been carried
out before. This analysis has been complemented with low-resolution, long-slit spectra that have allowed us to
describe the spectral characteristics of the nebula and their CSs. The main
conclusions of this work can be summarized as follows.

Abell\,36 presents a point-symmetric elliptical morphology but the
spatio-kinematical analysis reveals that it consists of a spheroidal shell and
two bright point-symmetric arcs, attributable to bipolar, rotating outflows;
the collimated outflows seems to have bored parts of the spheroid. DeHt\,2
appears as an elliptical PN in direct images but our analysis strongly
suggests that it has formed through two different 
ejection events, with the last one being more collimated than a previous
ellipsoidal shell; evidence also exists in DeHt\,2 for collimated outflow-shell interaction.
RWT\,152 is a bipolar PN with an equatorial ring. The complex structures of Abell\,36 and
DeHt\,2 suggest that binary central stars may be involved in their formation. 

The nebular spectra of the three PNe indicate a high excitation but only 
Abell\,36 exhibits emission lines different from those due to Balmer, [O\,{\sc
  iii}], and He\,{\sc ii}. In DeHt\,2 and RWT\,152 the nebular spectra suggest a
possible deficiency in heavy elements. 

The spectra of the three CSs present narrow absorption lines, being the
He\,{\sc ii}\,$\lambda$4686 absorption particularly
prominent. These characteristics, and the published atmospheric parameters
strongly suggest a sdO nature for these CSs. Thus, these sdOs
 have most probably evolved through AGB and post-AGB phases. 

The number of sdOs surrounded by PNe is very scarce ($\sim 18$). This
  number could be biased by the intrinsic faintness of the associated PNe, and by the 
lack of a firm classification for many CS of PNe. Very deep images of more sdOs, and analysis of 
high quality CS spectra are necessary to identify new PN+sdO systems. On the other hand, if sdOs originate
  from low-mass progenitors, the non-detection of more PN+sdO systems could
  be, at least for a certain fraction of sdOs, a consequence of the
  dissipation of the nebula before being photoionized, due to the slow evolution of the CS. 

We have compared properties of the 
  more confident PN+sdO systems and found that most of them are relatively or 
  very evolved PNe, present collimated outflows or signs that collimated outflows
have been involved in their formation, host binary central stars, and are
observed at relatively high Galactic latitudes. These properties and other
published results reinforce the idea that the formation of complex PNe is related to
binary stars rather than to the progenitor mass. More studies of PN+sdO
systems could provide interesting information about formation of complex PNe
and sdO evolution.

\section*{Acknowledgments}

We thank our anonymous referee for his/her useful comments
that have improved the interpretation and discussion of
the data. This paper has been supported partially by grant AYA\,2011-24052 (AA, ES),
AYA\,2011-30228-C3-01 (LFM), and AYA\,2011-30147-C03-01 (RO)
of the Spanish MICINN, and by grant INCITE09\,312191PR (AA, AU, LFM) of Xunta de Galicia, all of them
partially funded by FEDER funds. Partial support by grant 12VI20 of the Universidad de Vigo is also acknowledged.
 LO acknowledges support by project PROMEP/103.5/12/3590. RV acknowledges 
support from UNAM-DGAPA-PAPIIT grant IN107914. 
Authors also acknowledge the staff at OAN-SPM (in particular to Mr. Gustavo Melgoza-Kennedy), 
Calar Alto, and Roque de los Muchachos Observatories for support during observations.
 We acknowledge support from the 
Faculty of the European Space Astronomy Centre (ESAC). This
research has made use of the SIMBAD database, operated at the 
CDS, Strasbourg (France), Aladin, NASA's Astrophysics Data System
Bibliographic Services, and the Spanish Virtual Observatory supported from the
Spanish MEC through grant AYA2008-02156.

%\appendix

%\section[]{Large gaps in L\lowercase{y}${\balpha}$ forests\\* due to fluctuations in line distribution}

%This means that the Ly$\alpha$ forest lines are uniformly
%distributed in~$x$. The probability of finding $N-1$ lines between $z_1$
%and~$z_2$, $P_{N-1}$, is assumed to be the Poisson distribution.

%\begin{figure}
%\vspace{11pc}
%\caption{$P(>x_{\rmn{gap}})$ as a function of $x_{\rmn{gap}}$ for,
 %from left to right, $N=160$, 150, 140, 110, 100, 90, 50, 45 and~40.
% Compare this with \protect\citet{b15}.}
%\label{appenfig}
%\end{figure}

%\subsection{Subsection title}

%We plot in Fig.~\ref{appenfig} $P(>x_{\rmn{gap}})$ for several $N$
%values. We see that, for $N=100$ and $x_{\rmn{gap}}=0.06$,
%$P(>0.06)\approx20$ per cent.  This means that the probability of
%finding a gap with a size larger than six times the mean
%separation is not significantly small. 

%\bsp

\label{lastpage}


\begin{thebibliography}{99}

\bibitem[\protect\citeauthoryear{Abell}{1966}]{1966ApJ...144..259A} Abell 
G.~O., 1966, ApJ, 144, 259 

\bibitem[\protect\citeauthoryear{Acker et al.}{1992}]{1992secg.book.....A} 
Acker A., Marcout J., Ochsenbein F., Stenholm B., Tylenda R., Schohn C., 
1992, secg.book, 

\bibitem[\protect\citeauthoryear{Acker et 
al.}{1998}]{1998A&A...337..253A} Acker A., Fresneau A., Pottasch S.~R., Jasniewicz G., 1998, A\&A, 337, 253

\bibitem[\protect\citeauthoryear{Aller et 
al.}{2013}]{2013A&A...552A..25A} Aller A., et al., 2013, A\&A, 552, A25

\bibitem[\protect\citeauthoryear{Alves, Bond, 
\& Livio}{2000}]{2000AJ....120.2044A} Alves D.~R., Bond H.~E., Livio M., 2000, AJ, 120, 2044 

\bibitem[\protect\citeauthoryear{Boffin et al.}{2012}]{2012Sci...338..773B} 
Boffin H.~M.~J., Miszalski B., Rauch T., Jones D., Corradi R.~L.~M., 
Napiwotzki R., Day-Jones A.~C., K{\"o}ppen J., 2012, Sci, 338, 773 

\bibitem[\protect\citeauthoryear{Bohuski 
\& Smith}{1974}]{1974ApJ...193..197B} Bohuski T.~J., Smith M.~G., 1974, ApJ, 193, 197

\bibitem[\protect\citeauthoryear{Bond 
\& Livio}{1990}]{1990ApJ...355..568B} Bond H.~E., Livio M., 1990, ApJ, 355, 568

\bibitem[\protect\citeauthoryear{Bobrowsky et 
al.}{1998}]{1998Natur.392..469B} Bobrowsky M., Sahu K.~C., Parthasarathy 
M., Garc{\'{\i}}a-Lario P., 1998, Natur, 392, 469

\bibitem[\protect\citeauthoryear{Borkowski, Tsvetanov, 
\& Harrington}{1993}]{1993ApJ...402L..57B} Borkowski K.~J., Tsvetanov Z., Harrington J.~P., 1993, ApJ, 402, L57

\bibitem[\protect\citeauthoryear{Brocklehurst}{1971}]{1971MNRAS.153..471B} 
Brocklehurst M., 1971, MNRAS, 153, 471 

\bibitem[\protect\citeauthoryear{Bruch, Vaz, 
\& Diaz}{2001}]{2001A&A...377..898B} Bruch A., Vaz L.~P.~R., Diaz M.~P., 2001, A\&A, 377, 898

\bibitem[\protect\citeauthoryear{Cahn 
\& Kaler}{1971}]{1971ApJS...22..319C} Cahn J.~H., Kaler J.~B., 1971, ApJS, 22, 319 

\bibitem[\protect\citeauthoryear{Cahn, Kaler, 
\& Stanghellini}{1992}]{1992A&AS...94..399C} Cahn J.~H., Kaler J.~B., Stanghellini L., 1992, A\&AS, 94, 399 

\bibitem[\protect\citeauthoryear{Chromey}{1980}]{1980AJ.....85..853C} 
Chromey F.~R., 1980, AJ, 85, 853 

\bibitem[\protect\citeauthoryear{Chu et al.}{2009}]{2009AJ....138..691C} 
Chu Y.-H., et al., 2009, AJ, 138, 691

\bibitem[\protect\citeauthoryear{Ciardullo et 
al.}{1999}]{1999AJ....118..488C} Ciardullo R., Bond H.~E., Sipior M.~S., 
Fullton L.~K., Zhang C.-Y., Schaefer K.~G., 1999, AJ, 118, 488

\bibitem[\protect\citeauthoryear{Corradi 
\& Schwarz}{1993}]{1993A&A...273..247C} Corradi R.~L.~M., Schwarz H.~E., 1993, A\&A, 273, 247

\bibitem[\protect\citeauthoryear{Corradi et 
al.}{2014}]{2014MNRAS.441.2799C} Corradi R.~L.~M., et al., 2014, MNRAS, 
441, 2799 

\bibitem[\protect\citeauthoryear{Dengel, Hartl, 
\& Weinberger}{1980}]{1980A&A....85..356D} Dengel J., Hartl H., Weinberger R., 1980, A\&A, 85, 356 

\bibitem[\protect\citeauthoryear{De Marco}{2009}]{2009PASP..121..316D} De 
Marco O., 2009, PASP, 121, 316 

\bibitem[\protect\citeauthoryear{Drilling}{1983}]{1983ApJ...270L..13D} 
Drilling J.~S., 1983, ApJ, 270, L13

\bibitem[\protect\citeauthoryear{Ebbets 
\& Savage}{1982}]{1982ApJ...262..234E} Ebbets D.~C., Savage B.~D., 1982, ApJ, 262, 234

\bibitem[\protect\citeauthoryear{Frew et al.}{2014}]{2014arXiv1403.7847F} 
Frew D.~J., Bento J., Bojicic I.~S., Parker Q.~A., 2014, arXiv, 
arXiv:1403.7847 

\bibitem[\protect\citeauthoryear{Geier et al.}{2011}]{2011AIPC.1331..163G} 
Geier S., et al., 2011, AIPC, 1331, 163 

\bibitem[\protect\citeauthoryear{Geier}{2013}]{2013EPJWC..4304001G} Geier 
S., 2013, EPJWC, 43, 4001 

\bibitem[\protect\citeauthoryear{Goldman et 
al.}{2004}]{2004AJ....128.1711G} Goldman D.~B., Guerrero M.~A., Chu Y.-H., 
Gruendl R.~A., 2004, AJ, 128, 1711

\bibitem[\protect\citeauthoryear{Graham et al.}{2004}]{2004MNRAS.347.1370G} 
Graham M.~F., Meaburn J., L{\'o}pez J.~A., Harman D.~J., Holloway A.~J., 
2004, MNRAS, 347, 1370 

\bibitem[\protect\citeauthoryear{Grauer et al.}{1987}]{1987BAAS...19..643G} 
Grauer A.~D., Bond H.~E., Ciardullo R., Fleming T.~A., 1987, BAAS, 19, 643 

\bibitem[\protect\citeauthoryear{Guerrero et al.}{2008}]{2008ApJ...683..272G}
  Guerrero M.~A., et al., 2008, ApJ, 683, 272 

\bibitem[\protect\citeauthoryear{Guerrero \& Miranda}{2012}]{2012A&A...539A..47G} Guerrero M.~A., 
Miranda, L.~F., 2012, A\&A, 593, A47

\bibitem[\protect\citeauthoryear{Guill{\'e}n et 
al.}{2013}]{2013MNRAS.432.2676G} Guill{\'e}n P.~F., V{\'a}zquez R., Miranda 
L.~F., Zavala S., Contreras M.~E., Ayala S., Ortiz-Ambriz A., 2013, MNRAS, 
432, 2676 

\bibitem[\protect\citeauthoryear{Heber, Werner, 
\& Drilling}{1988}]{1988A&A...194..223H} Heber U., Werner K., Drilling J.~S., 1988, A\&A, 194, 223 

\bibitem[\protect\citeauthoryear{Heber}{2009}]{2009ARA&A..47..211H} Heber U., 2009, ARA\&A, 47, 211 

\bibitem[\protect\citeauthoryear{Heckathorn, Fesen, 
\& Gull}{1982}]{1982A&A...114..414H} Heckathorn J.~N., Fesen R.~A., Gull T.~R., 1982, A\&A, 114, 414

\bibitem[\protect\citeauthoryear{Herald 
\& Bianchi}{2004}]{2004ApJ...609..378H} Herald J.~E., Bianchi L., 2004, ApJ, 609, 378

\bibitem[\protect\citeauthoryear{Herrero, Manchado, 
\& Mendez}{1990}]{1990Ap&SS.169..183H} Herrero A., Manchado A., Mendez R.~H., 1990, Ap\&SS, 169, 183 


\bibitem[\protect\citeauthoryear{Hillwig et 
al.}{2010}]{2010AJ....140..319H} Hillwig T.~C., Bond H.~E., Af{\c s}ar M., 
De Marco O., 2010, AJ, 140, 319 

\bibitem[\protect\citeauthoryear{Hua, Dopita, 
\& Martinis}{1998}]{1998A&AS..133..361H} Hua C.~T., Dopita M.~A., Martinis J., 1998, A\&AS, 133, 361

\bibitem[\protect\citeauthoryear{Hua 
\& Kwok}{1999}]{1999A&AS..138..275H} Hua C.~T., Kwok S., 1999, A\&AS, 138, 275 

\bibitem[\protect\citeauthoryear{Jacoby et al.}{2012}]{2012AAS...21941802J} 
Jacoby G., De Marco O., Howell S., Kronberger M., 2012, AAS, 219, \#418.02 

\bibitem[\protect\citeauthoryear{Jones et al.}{2010}]{2010MNRAS.408.2312J} 
Jones D., et al., 2010, MNRAS, 408, 2312 

\bibitem[\protect\citeauthoryear{Jones et 
al.}{2014}]{2014A&A...562A..89J} Jones D., Boffin H.~M.~J., Miszalski B., Wesson R., Corradi R.~L.~M., Tyndall A.~A., 2014, A\&A, 562, A89

\bibitem[\protect\citeauthoryear{Kilkenny et 
al.}{1997}]{1997MNRAS.287..867K} Kilkenny D., O'Donoghue D., Koen C., 
Stobie R.~S., Chen A., 1997, MNRAS, 287, 867 

\bibitem[\protect\citeauthoryear{Kohoutek 
\& Hekela}{1967}]{1967BAICz..18..203K} Kohoutek L., Hekela J., 1967, BAICz, 18, 203

\bibitem[\protect\citeauthoryear{Kohoutek 
\& Senkbeil}{1973}]{1973MSRSL...5..485K} Kohoutek L., Senkbeil G., 1973, MSRSL, 5, 485 

\bibitem[\protect\citeauthoryear{Latour et al.}{2013}]{2013ApJ...773...84L} 
Latour M., Fontaine G., Chayer P., Brassard P., 2013, ApJ, 773, 84 

\bibitem[\protect\citeauthoryear{Leone et al.}{2011}]{2011ApJ...731L..33L} 
Leone F., Mart{\'{\i}}nez Gonz{\'a}lez M.~J., Corradi R.~L.~M., Privitera 
G., Manso Sainz R., 2011, ApJ, 731, L33 

\bibitem[\protect\citeauthoryear{Manchado et 
al.}{1996}]{1996iacm.book.....M} Manchado A., Guerrero M.~A., Stanghellini 
L., Serra-Ricart M., 1996, iacm.book, 

\bibitem[\protect\citeauthoryear{Meaburn et 
al.}{2003}]{2003RMxAA..39..185M} Meaburn J., L{\'o}pez J.~A., Guti{\'e}rrez 
L., Quir{\'o}z F., Murillo J.~M., Vald{\'e}z J., Pedrayez M., 2003, RMxAA, 
39, 185 

\bibitem[\protect\citeauthoryear{M{\'e}ndez}{1978}]{1978MNRAS.185..647M} 
M{\'e}ndez R.~H., 1978, MNRAS, 185, 647

\bibitem[\protect\citeauthoryear{M{\'e}ndez et 
al.}{1988}]{1988A&A...198..287M} Mendez R.~H., Gathier R., Simon K.~P., Kwitter K.~B., 1988, A\&A, 198, 287

\bibitem[\protect\citeauthoryear{M{\'e}ndez}{1991}]{1991IAUS..145..375M} M\'endez 
R.~H., 1991, IAUS, 145, 375 

\bibitem[\protect\citeauthoryear{Miranda 
\& Solf}{1992}]{1992A&A...260..397M} Miranda L.~F., Solf J., 1992, A\&A, 260, 397

\bibitem[\protect\citeauthoryear{Miszalski et 
al.}{2009}]{2009A&A...505..249M} Miszalski B., Acker A., Parker Q.~A., Moffat A.~F.~J., 2009, A\&A, 505, 249 

\bibitem[\protect\citeauthoryear{Mitchell et 
al.}{2007}]{2007MNRAS.374.1404M} Mitchell D.~L., Pollacco D., O'Brien 
T.~J., Bryce M., L{\'o}pez J.~A., Meaburn J., Vaytet N.~M.~H., 2007, MNRAS, 
374, 1404 

\bibitem[\protect\citeauthoryear{Monteiro et 
al.}{2000}]{2000ApJ...537..853M} Monteiro H., Morisset C., Gruenwald R., 
Viegas S.~M., 2000, ApJ, 537, 853 

\bibitem[\protect\citeauthoryear{Montez et al.}{2010}]{2010ApJ...721.1820M} 
Montez R., Jr., De Marco O., Kastner J.~H., Chu Y.-H., 2010, ApJ, 721, 1820

\bibitem[\protect\citeauthoryear{Nakanishi 
\& Sofue}{2003}]{2003PASJ...55..191N} Nakanishi H., Sofue Y., 2003, PASJ, 55, 191 

\bibitem[\protect\citeauthoryear{Napiwotzki}{1999}]{1999A&A...350..101N} Napiwotzki R., 1999, A\&A, 350, 101 

\bibitem[\protect\citeauthoryear{Napiwotzki}{2001}]{Napiwotzki_2001} Napiwotzki R., 2001, A\&A, 367, 973 

\bibitem[\protect\citeauthoryear{Napiwotzki}{2008}]{2008ASPC..392..139N} 
Napiwotzki R., 2008, ASPC, 392, 139 

\bibitem[\protect\citeauthoryear{Napiwotzki 
\& Schoenberner}{1995}]{1995A&A...301..545N} Napiwotzki R., Schoenberner D., 1995, A\&A, 301, 545 

\bibitem[\protect\citeauthoryear{{\O}stensen}{2006}]{2006BaltA..15...85O} 
{\O}stensen R.~H., 2006, BaltA, 15, 85 

\bibitem[\protect\citeauthoryear{Parthasarathy et 
al.}{1995}]{1995A&A...300L..25P} Parthasarathy M., et al., 1995, A\&A, 300, L25

\bibitem[\protect\citeauthoryear{Phillips}{2005}]{2005MNRAS.357..619P} 
Phillips J.~P., 2005, MNRAS, 357, 619 

\bibitem[\protect\citeauthoryear{Pollacco 
\& Bell}{1997}]{1997MNRAS.284...32P} Pollacco D.~L., Bell S.~A., 1997, MNRAS, 284, 32 

\bibitem[\protect\citeauthoryear{Pritchet}{1984}]{1984A&A...139..230P} Pritchet C., 1984, A\&A, 139, 230  

\bibitem[\protect\citeauthoryear{Ramos-Larios et 
al.}{2012}]{2012MNRAS.420.1977R} Ramos-Larios G., Guerrero M.~A., 
V{\'a}zquez R., Phillips J.~P., 2012, MNRAS, 420, 1977 

\bibitem[\protect\citeauthoryear{Rauch, Dreizler, 
\& Wolff}{1998}]{1998A&A...338..651R} Rauch T., Dreizler S., Wolff B., 1998, A\&A, 338, 651 

\bibitem[\protect\citeauthoryear{Reindl et 
al.}{2014}]{2014A&A...565A..40R} Reindl N., Rauch T., Parthasarathy M., Werner K., Kruk J.~W., Hamann W.-R., Sander A., Todt H., 2014, A\&A, 565, A40

\bibitem[\protect\citeauthoryear{Ritter 
\& Kolb}{2003}]{2003A&A...404..301R} Ritter H., Kolb U., 2003, A\&A, 404, 301

\bibitem[\protect\citeauthoryear{Seaton}{1979}]{1979MNRAS.187P..73S} Seaton 
M.~J., 1979, MNRAS, 187, 73P 

\bibitem[\protect\citeauthoryear{Shimanskii et 
al.}{2008}]{2008ARep...52..479S} Shimanskii V.~V., Borisov N.~V., 
Sakhibullin N.~A., Sheveleva D.~V., 2008, ARep, 52, 479 

\bibitem[\protect\citeauthoryear{Soker, Rappaport, 
\& Harpaz}{1998}]{1998ApJ...496..842S} Soker N., Rappaport S., Harpaz A., 1998, ApJ, 496, 842

\bibitem[\protect\citeauthoryear{Stanghellini et 
al.}{2002}]{2002ApJ...576..285S} Stanghellini L., Villaver E., Manchado A., 
Guerrero M.~A., 2002, ApJ, 576, 285 

\bibitem[\protect\citeauthoryear{Tocknell, De Marco \& Wardle}{2014}]{2014MNRAS.439.2014T} 
Tocknell J., De Marco O., Wardle M., 2014, MNRAS, 439, 2014

\bibitem[\protect\citeauthoryear{Traulsen et 
al.}{2005}]{2005ASPC..334..325T} Traulsen I., Hoffmann A.~I.~D., Rauch T., 
Werner K., Dreizler S., Kruk J.~W., 2005, ASPC, 334, 325

\bibitem[\protect\citeauthoryear{Tsessevich}{1977}]{1977IBVS.1320....1T} 
Tsessevich V.~P., 1977, IBVS, 1320, 1 

\bibitem[\protect\citeauthoryear{Van Winckel et 
al.}{2014}]{2014A&A...563L..10V} Van Winckel H., Jorissen A., Exter K., Raskin G., Prins S., Perez Padilla J., Merges F., Pessemier W., 2014, A\&A, 563, L10

\bibitem[\protect\citeauthoryear{V{\'a}zquez et 
al.}{2002}]{2002ApJ...576..860V} V{\'a}zquez R., Miranda L.~F., Torrelles 
J.~M., Olgu{\'{\i}}n L., Ben{\'{\i}}tez G., Rodr{\'{\i}}guez L.~F., 
L{\'o}pez J.~A., 2002, ApJ, 576, 860 

\bibitem[\protect\citeauthoryear{V{\'a}zquez et 
al.}{2008}]{2008A&A...481..107V} V{\'a}zquez R., Miranda L.~F., Olgu{\'{\i}}n L., Ayala S., Torrelles J.~M., Contreras M.~E., Guill{\'e}n P.~F., 2008, A\&A, 481, 107 

\bibitem[\protect\citeauthoryear{Wareing_etal}{2007}]{2007MNRAS.382.1233W} Wareing C.~J., Zijlstra A.~A., O'Brien T.~J., 2007, MNRAS, 382, 1233

\bibitem[\protect\citeauthoryear{Weidmann 
\& Gamen}{2011}]{2011A&A...526A...6W} Weidmann W.~A., Gamen R., 2011, A\&A, 526, A6

\bibitem[\protect\citeauthoryear{Wlodarczyk 
\& Olszewski}{1994}]{1994AcA....44..407W} Wlodarczyk K., Olszewski P., 1994, AcA, 44, 407 



\end{thebibliography}
\end{document}